\documentclass[10pt]{iopart}
\usepackage{iopams}
\usepackage{dsfont}
\usepackage{graphicx}
\usepackage{enumerate}
\graphicspath{{./figures/}}
\usepackage[square,sort&compress,numbers]{natbib}
\newcommand{\mri}{\mathrm{i}}
\newcommand{\Exp}[1]{\mathrm{e}^{\mbox{\footnotesize$#1$}}}

\renewcommand{\vec}[1]{\mathbf{#1}}

\begin{document}
\title{Non-equilibrium Atomic Condensates and Mixtures: Collective Modes, Condensate Growth and Thermalization}
\author{Kean Loon Lee and Nick P. Proukakis}
\address{Joint Quantum Centre (JQC) Durham-Newcastle, School of Mathematics and Statistics,
Newcastle University, Newcastle upon Tyne NE1 7RU, England, UK}
\eads{\mailto{keanloon.lee@ncl.ac.uk},\mailto{nikolaos.proukakis@newcastle.ac.uk}}
\begin{abstract}\noindent
The non-equilibrium dynamics of trapped ultracold atomic gases, or mixtures thereof, is an extremely rich subject. Despite 20 years of studies, and remarkable progress mainly on the experimental front, numerous open question remain, related to the growth, relaxation and thermalisation of such systems, and there is still no universally-accepted theory for their theoretical description. In this paper we discuss one of the state-of-the-art kinetic approaches, which gives an intuitive picture of the physical processes happening at the microscopic scale, being broadly applicable both  below and above the critical region (but not within the critical region itself, where fluctuations become dominant and symmetry breaking takes place).
Specifically, the ``Zaremba-Nikuni-Griffin" (ZNG) scheme provides a self-consistent description of the coupling between the condensate and the thermal atoms, including the collisions between these two subsystems.
It has been successfully tested against experiments in various settings, including investigation of collective modes (e.g. monopole, dipole and quadrupole modes), dissipation of topological excitations (solitons and vortices) as well as surface evaporative cooling. Here, we show that the ZNG model can capture two important aspects of non-equilibrium dynamics for both single-component and two-component BECs: the Kohn mode (the undamped dipole oscillation independent of interactions and temperature) and (re)thermalization leading to condensate growth following sudden evaporation. Our simulations, performed in a spherically-symmetric trap reveal (i) an interesting two-stage dynamics and the emergence of a prominent monopole mode in the evaporative cooling of a single-component Bose gas, and (ii) the long thermalization time associated with the sympathetic cooling of a realistic two-component mixture. Related open questions arise about the mechanisms and the nature of thermalization in such systems, where further controlled experiments are needed for benchmarking.
\end{abstract}
%\submitto{\jpb}
\maketitle
\ioptwocol

\section{Introduction}
Bose-Einstein condensates (BECs)~\cite{anderson_ensher_1995,davis_mewes_1995a,bradley_sackett_95,bradley_sackett_97} are interesting systems to study the non-equilibrium dynamics of interacting quantum gases. The ability to experimentally control the many-body interactions~\cite{chin_grimm_2010}, the temperature of the system~\cite{davis_mewes_1995b}, the trapping potentials~\cite{grimm_weidemuller_2000,gaunt_schmidutz_2013} as well as the coherent coupling between different hyperfine levels~\cite{stamper-kurn_ueda_2013}, makes BECs an ideal system to probe fundamental problems. In addition, the increasingly precise manipulation also brings about a growing interest in the applications of BECs to create quantum devices, such as an atomic SQUID~\cite{ryu_blackburn_2013,yi-hsieh_kumar_2015} or a matter-wave interferometer~\cite{pritchard_dinkelaker_2012,muntinga_ahlers_2013,bell_glidden_2016}. 

A problem that carries both fundamental importance as well as practical interests concerns the impact of temperature on the non-equilibrium dynamics of a partially-condensed Bose gas. A realistic experimental system typically consists of a condensate coexisting with non-condensed particles, which are collectively termed the \emph{thermal cloud}, at a temperature $T$ below the critical value $T_c$. For $0<T\ll T_c$, the thermal cloud exerts negligible influence on the condensate, hence there is a relatively simple description of the condensate through the nonlinear Schr\"odinger equation, also known as the Gross-Pitaevskii equation. Nevertheless, as the temperature increases, leading to a larger fraction of the thermal cloud, the interaction between the condensate and the thermal cloud can produce interesting and observable physical effects, such as damping of collective modes~\cite{jin_ensher_1996,jackson_zaremba_2002c,stamper-kurn_miesner_1998,hutchinson_zaremba_97,olshanii_98,shenoy_ho_1998,bijlsma_stoof_99,alkhawaja_stoof_00,marago_hechenblaikner_2001,jackson_zaremba_2001,chevy_bretin_2002,jackson_zaremba_2002a,ferlaino_maddaloni_2002,morgan_rusch_03,morgan_05,arahata_nikuni_2008,delehaye_laurent_2015}, energy dissipation of topological excitations~\cite{burger_bongs_1999,jackson_barenghi_2007,jackson_proukakis_2007,jackson_proukakis_2009,rooney_bradley_10,cockburn_nistazakis_10,cockburn_nistazakis_11,allen_zaremba_2013,allen_zuccher_2014,moon_kwon_2015,rooney_allen_2016}, defect formation~\cite{weiler_neely_08,lamporesi_donadello_2013,trento_prl,corman_chomaz_14,chomaz_corman_15,navon_gaunt_2015,liu_pattinson_2016} and thermalization~\cite{kagan_svistunov_1992,miesner_stamper-kurn_1998,gardiner_lee_1998,davis_gardiner_2000,bijlsma_zaremba_stoof_2000,svistunov_2001,harber_mcguirk_2003,proukakis_schmiedmayer_2006,gring_kuhnert_2012,handel_marchant_2012,markle_allen_2014}. Needless to say, understanding and controlling the decoherence effect of a quantum device at finite temperature are also of utmost importance in a practical setting.

Modelling the finite-temperature dynamics of a condensate is therefore of significant value. For a theoretical model to be useful, it should capture the essential physics with minimal input parameters. At the same time, it must also remain easy to use by solving either analytically or numerically with feasible amount of computational effort~\cite{proukakis_jackson_2008}. In this sense, the `Zaremba-Nikuni-Griffin' (ZNG) model~\cite{zaremba_nikuni_1999,nikuni_zaremba_1999,griffin_nikuni_2009} is very successful in modelling existing experiments on single-component Bose gases for a broad temperature range (typically $T/T_c < 0.9$, or $T > T_c$), with potential corrections at very low temperatues arising from the quasiparticle nature of the excitations. 
By construction, this model treats the interactions between the condensate and the thermal cloud fully self-consistently, so including all collisional processes between them, and the respective back-actions during their coupled dynamical evolution.
In that respect, it provides an intuitive picture of the dynamics happening at the microscopic scale and allows us to explore dynamics ranging from the collisionless evolution to the hydrodynamic regime. In particular, it takes into account three key aspects, which are not typically {\em simultaneously} accounted for {\em in full} by other finite-temperature models (such as `classical' or `c-field' methods) -- see e.g the reviews \cite{proukakis_gardiner_2013,proukakis_jackson_2008,berloff_brachet_14,griffin_nikuni_2009,blakie_bradley_08,brewczyk_gajda_07}: 
\begin{enumerate}
\item the dynamics of the thermal cloud (for example, this is only approximately included in so-called `classical field methods' \cite{blakie_bradley_08,proukakis_jackson_2008,rooney_blakie_12,proukakis_gardiner_2013,cockburn_proukakis_2009}); 
\item the spatially-dependent collisions and dissipation (the latter is typically ignored in stochastic treatments, but see e.g. the related stochastic soliton decay~\cite{cockburn_nistazakis_11}); and 
\item the conservation of the total number of atoms (note that this is guaranteed by construction here; for alternative explicitly number-conserving approaches, see Ref.~ \cite{gardiner_morgan_2014} and references therein).
\end{enumerate}

From a physics perspective, there are at least two stringent tests that a `good' finite temperature non-equilibrium theory should satisfy, namely the ability to
\begin{enumerate}
\item reproduce the Kohn mode, i.e. the undamped dipole oscillations for a shifted trap, an exact feature occuring irrespective of the interaction strength (this is essential for precision measurement of collective modes);
\item predict the full dynamical thermalization between the condensate and thermal cloud, achievable only through the fully self-consistent coupling of the dynamics of those two subsystems, and thus model the condensate growth process.
\end{enumerate}
The ZNG model satisfies both criteria, being valid both below and above the critical temperature (in the latter case it just reduces to the usual kinetic Boltzmann equation) \cite{griffin_nikuni_2009}. Here, it should be noted that, while the model does not account for the physics of the actual region of critical fluctuations close to $T_c$ (and so could not, for example, model phenomena associated with spontaneous defect formation {\em a la} Kibble-Zurek \cite{weiler_neely_08,lamporesi_donadello_2013,trento_prl,corman_chomaz_14,chomaz_corman_15,navon_gaunt_2015}) because of its explicit symmetry-breaking ansatz, it can nonetheless accurately predict condensate growth once a small condensate seed is added to the system \cite{bijlsma_zaremba_stoof_2000}, with the ensuing results being independent of the (small) seed size.
The interesting and related topic of condensate formation starting from an ultracold thermal bosonic system, has been thoroughly reviewed in \cite{davis_mybook}.

The extension of this ZNG model beyond the single-component Bose gas is rather involved. While the theoretical framework of the ZNG model has been established for a spinor condensate~\cite{nikuni_williams_2003,endo_nikuni_2011} (with coherent coupling between hyperfine levels) and two-component condensates~\cite{edmonds_lee_2015a,edmonds_lee_2015b} (with incoherent coupling between the two components), dynamical simulations of the two-component ZNG model have only been reported recently~\cite{lee_jorgensen_2016} and those for a spinor condensate remain unreported. Much work is still needed to perform systematic studies of two-component Bose gases, and to compare the experimental findings with the model predictions. 

The aim of this work is both to demonstrate that the ZNG model can capture the essential physics of non-equilibrium Bose gases at finite temperature, and also to use it to identify various deeper physical questions, where further experimental work is required to understand in detail. Such questions include the thermalization process during/after evaporative cooling, whether such a process involves just one, or more, timescales, the extent to which a binary mixture actually thermalizes on experimentally-relevant timescales, and more broadly the process of sympathetic cooling, for which we are not aware of any modelling describing the coupled dynamics once both components start exhibiting condensation.

After presenting a brief introduction of the ZNG model (\sref{sec:model}), including its recent extension to a two-component Bose-Einstein condensates~\cite{edmonds_lee_2015a,edmonds_lee_2015b}, we briefly explain how the model is solved numerically in practice in \sref{sec:zng_in_practice}. We then demonstrate the application of the model to study (i) the Kohn mode (\sref{sec:dipole}) and (ii) thermalization (\sref{sec:thermalization}), for both the single-component and the two-component cases. These examples serve as stringent tests of the validity of the ZNG model in capturing the essential physics.

\section{\label{sec:model}The Zaremba-Nikuni-Griffin (ZNG) Model}
The ZNG model was first obtained for a homogeneous Bose gas by Kirkpatrick and Dorfman~\cite{kirkpatrick_dorfman_1983,kirkpatrick_dorfman_1985}, and subsequently derived for an inhomogeneous single-component Bose gas~\cite{zaremba_nikuni_1999,nikuni_zaremba_1999,griffin_nikuni_2009}, an inhomogenous spinor gas~\cite{nikuni_williams_2003,endo_nikuni_2011} and an inhomogeneous two-component Bose-Bose mixture~\cite{edmonds_lee_2015a,edmonds_lee_2015b}. We present here a brief description of the two-component Bose gas model (which is reduced to the single-component Bose gas model if the inter-component interaction is switched off) in a way that it can also be easily understood by the non-experts.

We consider an interacting bosonic binary system described by the Hamiltonian
\begin{eqnarray}
\hat{H}=\int d{\bf r}\bigg\{\sum_{j}\hat{\Psi}^{\dagger}_{j}\bigg[-\frac{\hbar^2}{2m_j}\nabla^2+V_{j}({\bf r})\bigg]\hat{\Psi}_{j}\bigg\}+\hat{H}_{I}
\end{eqnarray}
and the two-body interactions are given by
\begin{eqnarray}
\hat{H}_{I}{=}{\int}{d{\bf r}}\bigg\{\sum_{j}\frac{g_{jj}}{2}\hat{\Psi}^{\dagger}_{j}\hat{\Psi}^{\dagger}_{j}\hat{\Psi}_{j}\hat{\Psi}_{j}{+}\sum_{k\neq j}g_{kj}\hat{\Psi}^{\dagger}_{j}\hat{\Psi}^{\dagger}_{k}\hat{\Psi}_{k}\hat{\Psi}_{j}\bigg\},\nonumber\\
\quad
\end{eqnarray}
where $\hat{\Psi}_{j}\equiv\hat{\Psi}_{j}({\bf r}) \left(\hat{\Psi}_{j}^\dag\equiv\hat{\Psi}_{j}^\dag({\bf r})\right)$ is the bosonic annihilation (creation) operator for an atom of component-$j$ with mass $m_j$, which obeys the usual commutation relationships for bosons, 
\begin{equation}
\eqalign{{[}\hat{\Psi}_j({\bf r}),\hat{\Psi}^{\dag}_{k}({\bf r}')]=\delta_{kj}\delta({\bf r}-{\bf r}'),\cr
{[}\hat{\Psi}_{j}({\bf r}),\hat{\Psi}_{k}({\bf r}')]=[\hat{\Psi}^{\dag}_{j}({\bf r}),\hat{\Psi}^{\dag}_{k}({\bf r}')]=0.}
\end{equation}
The $s$-wave collisions between atoms in different components are encompassed by $g_{kj}=2\pi\hbar^2 a_{kj}/m_{kj}$, where $a_{kj}$ defines the scattering length between atoms in components $j$ and $k$, and $m^{-1}_{kj}=m^{-1}_{j}+m^{-1}_{k}$ defines the reduced mass. The atoms are confined in a harmonic potential $V_j(\vec{r})=\frac{1}{2}m_j[\omega_{r,j}^2 r^2 + \omega_{z,j}^2 z^2]$ with radial and axial angular frequencies, $\omega_{r,j}$ and $\omega_{z,j}$, respectively. For simplicity, we also assume isotropic traps ($\omega_{r,j} = \omega_{z,j}=\omega_j$) in our analysis. Also, for subsequent discussions that involve only a single component, we omit the subscript $j$ to simplify the notation.

Central to the ZNG methodology is the symmetry-breaking ansatz, and the Beliaev decomposition of the Bose field operator $\hat{\Psi}_{j}$ into its average, non-zero, value $\phi_j = \langle\hat{\Psi}_{j}\rangle$ denoting the condensate wavefunction, and a fluctuation operator $\hat{\delta}_j$, where angular brackets $\langle\ldots\rangle$ denote the broken-symmetry ensemble average\footnote{The effect of making such a decomposition has been discussed extensively in the Chapter by Griffin and Zaremba in Ref.~\cite{proukakis_gardiner_2013}, and the associated chapter by Davis, Wright and Proukakis in the same book.}. It is precisely because of this decomposition (and the assumption of a non-zero expectation value for the Bose field operator) that the ZNG model cannot account for condensate growth from a purely thermal system (i.e. without introducing a numerically-convenient non-zero seed).
The condensate density $n_{c,j}$ and the thermal cloud density $\tilde{n}_j$ for atoms of component $j$ are then given separately by the wavefunction,
\begin{equation}
n_{c,j} = |\phi_j|^2
\end{equation}
and the fluctuation operator via the \emph{diagonal} noncondensate density
\begin{equation}
\tilde{n}_j = \langle \delta_j^\dag \delta_j\rangle.
\end{equation}

A kinetic model is developed in \cite{edmonds_lee_2015a,edmonds_lee_2015b}, in which we identify the condensate field $\phi_j(\vec{r}) = \langle \hat{\Psi}_j^{\dagger}(\vec{r})\rangle $ and the thermal cloud density as the only \emph{slowly-varying} relevant quantities. The triplet anomalous averages $\langle \hat{\delta}^\dag_j\hat{\delta}_j\hat{\delta}_j\rangle$ and $\langle \hat{\delta}^\dag_k\hat{\delta}_k\hat{\delta}_j\rangle$, as well as the \emph{off-diagonal} noncondensate density $\langle \hat{\delta}^\dag_k \hat{\delta}_j\rangle$ for $k\neq j$, are treated perturbatively via adiabatic elimination~\cite{proukakis_burnett_1996,proukakis_burnett_1998}. In addition, the pair anomalous averages $\langle \hat{\delta}_j \hat{\delta}_j\rangle$ and $\langle \hat{\delta}_k \hat{\delta}_j\rangle$ are discarded~\cite{griffin_1996} as they do not generate energy-conserving contributions (to order $g^2$). The end result is that a condensate field $\phi_j(\vec{r})$ obeys a dissipative Gross-Pitaevskii equation
\begin{equation}
i\hbar\frac{\partial\phi_j}{\partial t}=\bigg[-\frac{\hbar^2}{2m_j}\nabla^2+U^{j}_{c}-i(R^{jj}+R^{kj}+\mathds{R}^{kj})\bigg]\phi_j,\label{eq:dschro1}
\end{equation}
while the Wigner distribution function of the thermal atoms 
\begin{equation}
f^j(\vec{p},\vec{r},t) = \int d\vec{r}' \Exp{\mri \vec{p}\cdot\vec{r}'/\hbar}\langle \delta^\dag\left(\vec{r}+\frac{\vec{r}'}{2},t\right)\delta\left(\vec{r}-\frac{\vec{r}'}{2},t\right)\rangle
\end{equation}
obeys a quantum Boltzmann equation
\begin{eqnarray}\nonumber\label{eq:qbe}
&\frac{\partial}{\partial t}f^{j}+\frac{1}{m_j}{\bf p}\cdot\nabla_{\bf r}f^{j}-\nabla_{\bf p}f^{j}\cdot\nabla_{\bf r}U^{j}_{\rm n}\\
&=\bigg(C^{jj}_{12}+C^{kj}_{12}\bigg)+\mathds{C}^{kj}_{12}+\bigg(C^{jj}_{22}+C^{kj}_{22}\bigg).
\end{eqnarray}
Compared to the usual zero-temperature case, the condensate atoms now experience an effective potential $U_c^j$ corrected by the presence of the thermal cloud, while the thermal atoms are modelled as classical particles moving in an effective potential $U_n^j$. These potentials include both the external potential $V_j$ and the mean-field contributions, which are related to the condensate density $n_{c,j}(\vec{r}) = |\phi_j(\vec{r})|^2$, and the thermal cloud density $\tilde{n}_j(\vec{r}) = \int d\vec{p}/(2\pi\hbar)^3 f^j(\vec{p},\vec{r},t)$, as
\begin{numparts}
\begin{eqnarray}
\label{eq:effUc}U_c^j =& V_j+g_{jj}(n_{c,j} +2\tilde{n}_j)+g_{kj}(n_{c,k} +\tilde{n}_k),\\
\label{eq:effUt}U_n^j =& V_j+ 2g_{jj}(n_{c,j} + \tilde{n}_j) +g_{kj}(n_{c,k} + \tilde{n}_k).
\end{eqnarray}
\end{numparts}
With these effective potentials, locally a condensate atom of component $j$ has energy
\begin{equation}
\varepsilon_c^j = \mu_c^j + \frac{1}{2}m_j\vec{v}_{c,j}^2,\label{eq:conEnergy}
\end{equation}
where 
\begin{equation}
\mu_{c}^{j}=-\frac{\hbar^2}{2m_j\sqrt{n_{c,j}}}(\nabla^2\sqrt{n_{c,j}})+U^{j}_{c}
\end{equation}
is the non-equilibrium chemical potential for component $j$ and 
\begin{equation}
{\bf v}_{c,j}=\frac{\hbar}{m_j n_{c,j}}{\rm Im} (\phi_j^*\nabla\phi_j)
\end{equation} 
defines the superfluid velocity of component $j$ with momentum $\vec{p}_c^j = m_j \vec{v}_{c,j}$. On the other hand, a thermal atom of component $j$ with momentum $\vec{p}$ has a Hartree energy
\begin{equation}
\varepsilon_{\vec{p}}^j = \frac{\vec{p}^2}{2m_j} + U_n^j.
\end{equation}
\Eref{eq:effUc} and \eref{eq:effUt} encapsulate the mean-field effects between the condensates and the thermal clouds, as well as the mean-field effects between the different components. As the condensates and the thermal clouds evolve in time, the changes in densities cause the clouds to exert a force on each other through the mean-field potentials and equations \eref{eq:dschro1} and \eref{eq:qbe}, leading to a damping effect, without explicit consideration of collisions. As a consequence, calculating equations \eref{eq:effUc} and \eref{eq:effUt} dynamically allows us to simulate the collisionless evolution and study the \emph{Landau damping}~\cite{jackson_zaremba_2003}.

The collisional integrals $C_{22}^{\cdot\cdot}$, $C_{12}^{\cdot\cdot}$ and $\mathds{C}_{12}^{kj}$ in equation \eref{eq:qbe} are important to establish the full thermal equilibrium of the system starting from a non-equilibrium state, and they are responsible for the \emph{collisional damping}. In particular, the thermal-thermal collisional integral,
\begin{eqnarray}\label{c22jk}
C_{22}^{kj} = &{\frac{g_{kj}^2}{(2\pi)^5\hbar^7}}(1+\delta_{kj}){\int{d{\bf p}_2}}{\int d{\bf p}_3}{\int d{\bf p}_4}\\
&\times\delta({\bf p}{+}{\bf p}_2{-}{\bf p}_3{-}{\bf p}_4){\delta(\varepsilon^{j}_{\vec{p}}{+}\varepsilon^{k}_{\vec{p}_2}{-}\varepsilon^{k}_{\vec{p}_3}{-}\varepsilon^{j}_{\vec{p}_4})}\nonumber\\
&\times\bigg[{(f^{j}{+}1)(f^{k}_{2}{+}1)f^{k}_{3}f^{j}_{4}{-}f^{j}f^{k}_{2}(f^{k}_{3}{+}1)(f^{j}_{4}{+}1)}\bigg],\nonumber
\end{eqnarray}
between thermal atoms of component $k$ and component $j$ (including both $k=j$ and $k\neq j$) vanishes when the thermal atoms are in local thermodynamic equilibrium,
\begin{equation}
f^j(\vec{p},\vec{r},t) = \{\Exp{\beta[(\vec{p}-m_j\vec{v}_{n,j})^2/2m_j + U_n^j - \tilde{\mu}_j]}-1\}^{-1},
\end{equation}
with the inverse temperature $\beta = (k_B T)^{-1}$, the normal fluid velocity $\vec{v}_{n,j}$ and the normal fluid chemical potential $\tilde{\mu}_j$, all as a function of position $\vec{r}$ and time $t$. Here, $k_B$ is the Boltzmann constant.

On the other hand, the thermal-condensate collisional integral,
\begin{eqnarray}\label{c12jk}\nonumber
C^{kj}_{12}=&\frac{g_{kj}^{2}}{(2\pi)^2\hbar^4}(1+\delta_{kj})n_{c,k}\int d{\bf p}_{2}\int d{\bf p}_{3}\int d{\bf p}_{4}\\\nonumber
&\times\delta({\bf p}_{c}^{k}+{\bf p}_{2}-{\bf p}_{3}-{\bf p}_{4})\delta(\varepsilon^{k}_{c}+\varepsilon^{j}_{\vec{p}_2}-\varepsilon^{j}_{\vec{p}_3}-\varepsilon^{k}_{\vec{p}_4})\\\nonumber
&\times\bigg[(f^{j}_{2}+1)f^{j}_{3}f^{k}_{4}-f^{j}_{2}(f^{j}_{3}+1)(f^{k}_{4}+1)\bigg]\\\nonumber
&\times\bigg[\delta({\bf p}-{\bf p}_{2})-\delta({\bf p}-{\bf p}_{3})\bigg]\\\nonumber
-&\frac{g_{kj}^{2}}{(2\pi)^2\hbar^4}(1+\delta_{kj})n_{c,j}\int d{\bf p}_{2}\int d{\bf p}_{3}\int d{\bf p}_{4}\\\nonumber
&\times\delta({\bf p}_{c}^{j}+{\bf p}_{2}-{\bf p}_{3}-{\bf p}_{4})\delta(\varepsilon^{j}_{c}+\varepsilon^{k}_{\vec{p}_2}-\varepsilon^{k}_{\vec{p}_3}-\varepsilon^{j}_{\vec{p}_4})
\\\nonumber
&\times\bigg[(f^{k}_{2}+1)f^{k}_{3}f^{j}_{4}-f^{k}_{2}(f^{k}_{3}+1)(f^{j}_{4}+1)\bigg]\nonumber\\
&\times\delta({\bf p}-{\bf p}_4),
\end{eqnarray}
 (including both $k=j$ and $k\neq j$) leads to a change in the number of condensate atoms of component $j$ through the source term 
 \begin{equation}\label{eq:r12}
 R^{kj}=\frac{\hbar}{2n_{c,j}}\int \frac{d\vec{p}}{(2\pi\hbar)^3}C_{12}^{kj},
 \end{equation}
 such that the total number of atoms (condensate + thermal cloud) remains unchanged. The $C^{kj}_{12}$ integral, and consequently the source term $R^{kj}$, only vanishes when the condensate atoms and the thermal atoms reach a local diffusive equilibrium (i.e. $\mu_{c}^j$ = $\tilde{\mu}_j$)~\cite{nikuni_zaremba_1999,nikuni_griffin_2001}.
 
In contrast to a spinor BEC, a two-component mixture admits a condensate-exchange collision that comes from a perturbative treatment of the normal pair average $\langle \hat{\delta}^\dag_k \hat{\delta}_j\rangle$. The corresponding collisional integral
\begin{eqnarray}
\mathds{C}^{kj}_{12}=&\frac{2\pi g_{kj}^{2}}{\hbar}\ n_{c,k}\,n_{c,j}\,\int d{\bf p}_{1}\int d{\bf p}_{2}\nonumber\\
&\times\delta({\bf p}_{c}^{j}+{\bf p}_{1}-{\bf p}_{c}^{k}-{\bf p}_{2})\delta(\varepsilon^{j}_{c}+\varepsilon^{k}_{p_1}-\varepsilon^{k}_{c}-\varepsilon^{j}_{p_2}
)\nonumber\\
&\times\bigg[f^{k}_{1}(f^{j}_{2}+1)-(f^{k}_{1}+1)f^{j}_{2}\bigg]\delta({\bf p}-{\bf p}_2).
\label{c12jk2}
\end{eqnarray}
for $k\neq j$ and the source term
\begin{equation}\label{eq:r_xc}
\mathds{R}^{kj}=\frac{\hbar}{2n_{c,j}}\int \frac{d\vec{p}}{(2\pi\hbar)^3}\mathds{C}_{12}^{kj}
\end{equation}
describe a collisional process whereby (in the forward process) one thermal atom of component $k$ collides with a condensate atom of component $j$ and promotes the condensate atom to the thermal cloud while itself being cooled and condenses into the condensate. When evaluated with respect to realistic mixtures at thermal equilibrium, this condensate-exchange collision has dominant collisional rates~\cite{edmonds_lee_2015a,edmonds_lee_2015b}, but its impact in dynamical situations remains a subject of study.
%The source terms in Eq.~\eqref{eq:dschro1} are related to the collision integrals in Eq.~\eqref{eq:qbe} via $R^{kj}({\bf r},t)=\frac{\hbar}{2n_{c,j}}\int\frac{d{\bf p}}{(2\pi\hbar)^3}C^{kj}_{12}$ (for $k=j$ and $k\neq j$) and $\mathds{R}^{kj}({\bf r},t)=\frac{\hbar}{2n_{c,j}}\int\frac{d{\bf p}}{(2\pi\hbar)^3}\mathds{C}^{kj}_{12}$. These collision integrals are sampled in dynamical simulation through direct simulation Monte Carlo~\cite{jackson_zaremba_2002}.

\section{\label{sec:zng_in_practice}Solving the ZNG model in practice}
Despite the two equations~\eref{eq:dschro1} and \eref{eq:qbe} that summarize the model being simple-looking, solving the equations for an arbitrary non-equilibrium situation is a numerically-challenging task and, to the best of our knowledge, very few groups have achieved it. Its first numerical implementation is carried out and documented by Jackson and Zaremba~\cite{jackson_zaremba_2002}. This code is subsequently adapted for the various studies being carried out at Newcastle~\cite{jackson_barenghi_2007,jackson_proukakis_2007,jackson_proukakis_2009,allen_zaremba_2013,allen_zuccher_2014,rooney_allen_2016}. Arahata and Nikuni reported its application to study first and second sound in a highly-elongated trap~\cite{arahata_nikuni_2013}. A parallel version sped up with OpenMPI has been developed by M\"arkle~\cite{markle_2014} to study surface evaporative cooling~\cite{markle_allen_2014}. We have recently developed a parallel version of the two-component ZNG model, for which we speed up the computation with OpenMP. We have used our new code to study the collective modes and rethermalization dynamics of both a single-component Bose gas and a two-component mixture, some of which are reported in this work. In particular, our recent application of the ZNG model~\cite{lee_jorgensen_2016} to study the counterflow dipole oscillation of a two-component mixture is the first of its kind and helps to identify the use of the dipole oscillation to map out the miscible-immiscible transition of a mixture. Currently, we are developing a new numerical code for use on an Nvidia graphics processing unit. Most recently, Straastma at JILA has exploited the spherical symmetry and implemented the ZNG approach for an isotropic trap to study the damping of monopole oscillation below the critical temperature~\cite{straastma_colussi_2016}.

The basic algorithm to solve the ZNG model is outlined in~\cite{jackson_zaremba_2002}. We first obtain the equilibrium distributions at a finite temperature $T$. In this case, both the source terms and the collision integrals vanish, hence we can set $\phi_j(\vec{r},t) = \phi_j(\vec{r})\Exp{-\mri\mu_j t/\hbar}$ with chemical potential $\mu_j$ and use a semi-classical Hartree-Fock approximation for the thermal cloud 
\begin{equation}\label{eq:eq_ntherm}
\tilde{n}_j(\vec{r})= g_{3/2}(z_j)/\lambda_j^3
\end{equation}
with thermal wavelength $\lambda_j = \sqrt{2\pi\hbar^2/(m_j k_B T)}$, local fugacity $z_j = \exp[(\mu_j-U_n^j )/(k_B T )]$ and the textbook result of Bose function,
\begin{equation}
g_{3/2}(z)=\frac{2}{\sqrt{\pi}} \int_0^\infty dx \frac{\sqrt{x}}{\Exp{x}/z-1}.
\end{equation}
We then obtain the condensate and thermal density profiles by solving equations
\eref{eq:dschro1} (through imaginary-time propagation) and \eref{eq:eq_ntherm} self-consistently. The equilibrium condensate wavefunctions and the thermal cloud densities are then set as the initial conditions of equations \eref{eq:dschro1} and \eref{eq:qbe} with appropriate modifications.

In a typical dynamical simulation, the dissipative Gross-Pitaevskii equation~\eref{eq:dschro1} is solved with the highly-efficient Fourier split-step method while the quantum Boltzmann equation~\eref{eq:qbe} is solved with the direct simulation Monte Carlo (DSMC) method~\cite{bird_1976,bird_1994}. A large number of test particles (typically of the order of millions) are generated according to the Bose-Einstein distribution $[\Exp{(\varepsilon^j_{\vec{p}}-\mu_j)/(k_BT)}-1]^{-1}$. These test particles are then evolved in time according to Newton's equation of motion using the symplectic leapfrog method. This solves the left-hand side of equation \eref{eq:qbe}.

In order to simulate the collisions and to calculate the source terms, i.e. solving the right-hand side of equation \eref{eq:qbe} and estimating equations \eref{eq:r12} and \eref{eq:r_xc}, the test particles are binned into spatial cells, where the cells have adaptive volumes~\cite{gallis_torczynski_2009,wade_baillie_2011} to improve the speed and accuracy of the computations. Collisions are simulated by randomly selecting pairs of test particles belonging to the same cell and checking if they are going to collide using the acceptance-rejection method. The exact forms of the probabilities used in the acceptance-rejection method have been given in~\cite{jackson_zaremba_2002,griffin_nikuni_2009}. It is important to note that the same probabilities are also used to compute the source terms \eref{eq:r12} and \eref{eq:r_xc}, hence accurate estimates of the source terms are accompanied by a large number of checks on collision events. In other words, while it is possible to speed up the simulation of the $C^{\cdot\cdot}_{22}$ collisions through the scaling of probabilities~\cite{bird_1994,wade_baillie_2011}, this is no longer feasible for the thermal-condensate collisions ($C^{\cdot\cdot}_{12}$ and $\mathds{C}^{kj}_{12}$ processes). From our experience, the quality of a full ZNG simulation that includes all collisional processes is largely determined by how well we have simulated the collisions and calculated the source terms.

In the next two sections, we will study the dipole oscillations and rethermalization of both a single-component Bose gas as well as a Bose-Bose mixture, by solving the full ZNG model, where all collision processes have been included, unless it is stated otherwise.

\section{\label{sec:dipole}Example 1: Dipole oscillation / Kohn mode}
A very stringent test of any finite-temperature model of an interacting Bose gas is the dipole mode, or more commonly known as the Kohn mode~\cite{kohn_1961,dobson_1994}. For a \emph{harmonically-trapped} gas, the center-of-mass (COM) degree of freedom is decoupled from all other internal degrees of freedom, hence the COM of a Bose gas would exhibit an \emph{undamped} dipole oscillation at the trap frequency, independent of the interactions among the particles or the temperature of the gas. In the presence of a thermal cloud at finite temperature, the Kohn mode manifests as the undamped in-phase oscillation of the condensate and the thermal cloud~\cite{stamper-kurn_miesner_1998}, while the out-of-phase oscillation displays damping and frequency shift~\cite{stamper-kurn_miesner_1998,meppelink_koller_2009,yuen_barr_2015}. The Kohn mode thus serves as a very accurate way to measure the trap frequency~\cite{dalfovo_giorgini_1999}. 

\begin{figure}
  \includegraphics[width=0.49\textwidth]{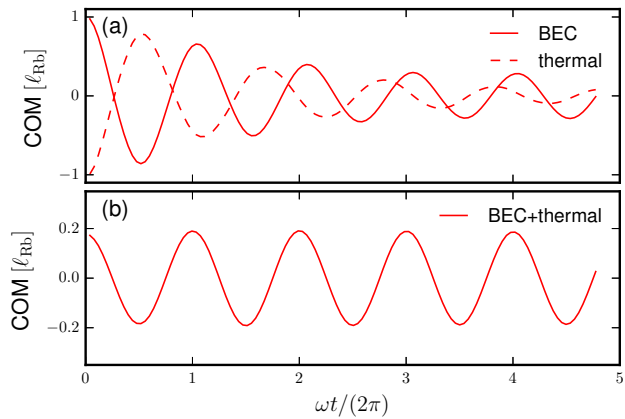}
  \caption{\label{fig:single_dipole} Dipole oscillation of $10^5$ $^{87}$Rb atoms at 250$\,$nK in an isotropic trap with angular trap frequency $\omega=2\pi\times200\,$Hz and scattering length $a_{\rm Rb}=99a_0$. The figure shows (a) the COMs of the condensate (red solid) and the thermal cloud (red dash) and (b) the COM of the whole cloud as a function of time $t$. See online supplementary video for the variation of density profiles with respect to time.}    
\end{figure}

\subsection{Kohn Mode for a Single Condensate}

In order to correctly reproduce the undamped Kohn mode for a single-component condensate, it is crucial to include the dynamics of both the condensate and the thermal cloud, and to couple them self-consistently. To see this, we can write down the coupled equations of motion of the COM along the $z$-axis as
\begin{numparts}
\begin{eqnarray}
\label{eq:eqm_con1}m N_c\frac{d^2z_c}{dt^2} = -m N_c\omega^2 z_c + F_{c,t},\\
\label{eq:eqm_therm1}m N_t\frac{d^2z_t}{dt^2} = -m N_t\omega^2 z_t + F_{t,c},
\end{eqnarray}
\end{numparts}
where
\begin{numparts}
\begin{eqnarray}
\label{eq:zc}z_c=\frac{1}{N_c} \int d\vec{r}\,z\,n_c(\vec{r}),\\
\label{eq:zt}z_t=\frac{1}{N_t}\int d\vec{r}\,z\,\tilde{n}(\vec{r}),
\end{eqnarray}
\end{numparts}
define the COMs of the condensate \eref{eq:zc} and the thermal cloud \eref{eq:zt} and 
\begin{eqnarray}
N_c = \int d\vec{r}\,n_c(\vec{r})\quad {\rm and}\quad N_t = \int d\vec{r}\,\tilde{n}(\vec{r})
\end{eqnarray}
give the number of condensate and thermal atoms respectively. $F_{c,t}$ is the force acting on the condensate due to the thermal cloud, and vice versa for $F_{t,c}$. The COM of the whole atomic cloud, including both the condensate and the thermal atoms, is stated in terms of
\begin{equation}
z_{\rm tot} = \frac{1}{N_c+N_t}\int d\vec{r}\,z\,[n_c(\vec{r}) + \tilde{n}(\vec{r})],
\end{equation}
and its equation of motion is obtained by simply adding up \eref{eq:eqm_con1} and \eref{eq:eqm_therm1}. 

Neglecting the thermal cloud dynamics [i.e. omitting \eref{eq:eqm_therm1}] or not coupling the condensate and the thermal cloud self-consistently (e.g. $F_{c,t}\neq -F_{t,c}$) would therefore not lead to the correct equation of motion for $z_{\rm tot}$,
\begin{equation}
\label{eq:eqm_tot1}\frac{d^2z_{\rm tot}}{dt^2} = -\omega^2 z_{\rm tot}.
\end{equation}

In \fref{fig:single_dipole}, we show that the ZNG model can correctly capture the Kohn mode described by equation \eref{eq:eqm_tot1}~\cite{jackson_zaremba_2002b}. We start with $10^5$ $^{87}$Rb atoms at thermal equilibrium (temperature $T=250\,$nK, 60\% condensate) in an isotropic trap with angular trap frequency $\omega=2\pi\times200\,$Hz. The scattering length is chosen to be $a_{\rm Rb}=99a_0$. At the beginning of our dynamical simulation, we rigidly shift the condensate and the thermal cloud by one harmonic trap length ($\ell_{\rm Rb}=0.76\mu$m) but in opposite directions along the $z$-axis. As the clouds oscillate out of phase, they exert a force on each other, causing damping in their oscillations~(\fref{fig:single_dipole}a)~\cite{stamper-kurn_miesner_1998,meppelink_koller_2009,yuen_barr_2015}. However, the total COM, shown in \fref{fig:single_dipole}b, displays an undamped oscillation that gives no hint of the internal dynamics. Moreover, the oscillation frequency corresponds precisely to the trap frequency.

\begin{figure}
  \includegraphics[width=0.49\textwidth]{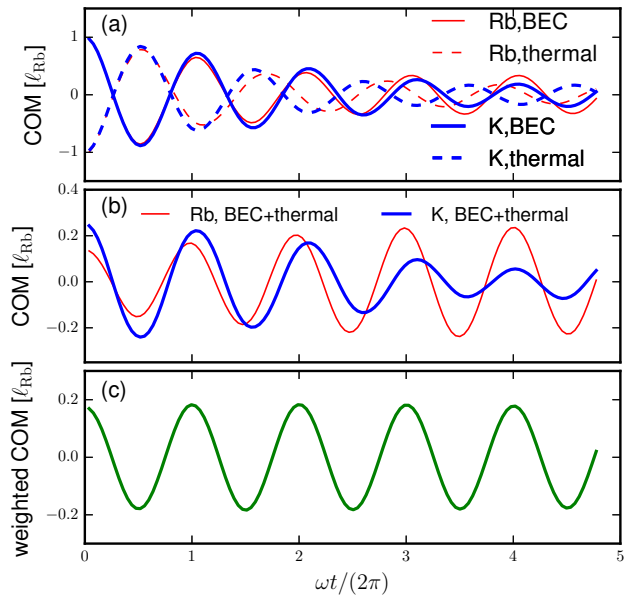}
  \caption{\label{fig:binary_dipole} Dipole oscillation of $^{87}$Rb (thin red lines) and $^{41}$K (thick blue lines) atoms, each with $10^5$ atoms at temperature 250$\,$nK and in isotropic traps with angular trap frequency $\omega=2\pi\times200\,$Hz, scattering lengths $a_{\rm Rb}=99a_0$, $a_{\rm K}=60 a_0$ and $a_{\rm Rb-K}=20a_0$~\cite{modugno_modugno_2002,thalhammer_barontini_2008}. The figure shows (a) the COMs of the condensates (solid) and the thermal clouds (dash), (b) the COM of the whole cloud for each component, and (c) the weighted COM of a two-component mixture~\eref{eq:weight_com}, as a function of time $t$. See online supplementary video for the variation of density profiles with respect to time.}    
\end{figure}

\subsection{Kohn Mode for a Binary Mixture}

For a two-component mixture, the dipole oscillation is very useful in measuring the miscible-immiscible transition~\cite{lee_jorgensen_2016}. In constrast to the single-component case, where a Kohn mode exists for arbitrary trap frequency, a two-component mixture would exhibit a Kohn mode only if $\omega_1 = \omega_2$. Similar to the single-component case, we can write down the equation of motion of the COM of the $i$th component,
\begin{equation}
\label{eq:eqm_binary}m_i N_i\frac{d^2z_{{\rm tot}, i}}{dt^2} = -m_i N_i\omega_i^2 z_{{\rm tot}, i} + F_{ij},
\end{equation}
where $F_{ij}$ is the sum of all forces acting on the $i$th component due to the $j$th component. A self-consistent model necessarily imposes the restriction $F_{ij}=-F_{ji}$, hence the weighted COM,
\begin{equation}
\label{eq:weight_com}z_{\rm weighted} = \sum_i m_i N_i z_{{\rm tot}, i} / \left(\sum_i m_i N_i\right)
\end{equation}
would oscillate at the trap frequency if $\omega_1 = \omega_2$. 

We demonstrate in \fref{fig:binary_dipole} that the ZNG model can capture the Kohn mode for a binary mixture. We have $^{87}$Rb and $^{41}$K with $10^5$ atoms each in isotropic traps with frequency $\omega_{\rm Rb}=\omega_{\rm K}=\omega=2\pi\times 200\,$Hz. The scattering lengths are $a_{\rm Rb}=99 a_0$, $a_{\rm K}=60 a_0$ and $a_{\rm Rb-K}=20a_0$~\cite{modugno_modugno_2002,thalhammer_barontini_2008}. The temperatures of both components are 250$\,$nK, which lead to 57\% and 63\% condensate fractions for $^{87}$Rb and $^{41}$K respectively. Similar to the single-component study, we shift the condensates and the thermal clouds by one harmonic trap length of $^{87}$Rb ($\ell_{\rm Rb}=0.76\mu$m) in opposite direction at the beginning of our dynamical simulation. The subsequent dynamics of the condensates (solid lines in \fref{fig:binary_dipole}a), the thermal clouds (dashed lines in \fref{fig:binary_dipole}b) as well as the individual components as a whole (\fref{fig:binary_dipole}b), are distinctly different from those of a simple harmonic oscillator. In constrast, the weighted COM (\fref{fig:binary_dipole}c) clearly behaves like a simple harmonic oscillator oscillating at the trap frequency, proving that the ZNG model can reproduce the Kohn mode for a two-component mixture.

To the best of our knowledge, this is the only numerically-viable approach which accurately models the Kohn mode at finite temperature. Classical field (e.g. `PGPE') \cite{bradley_blakie_2005}, or stochastic (e.g. SGPE / SPGPE) methods by construction violate this, due to the lack of a dynamical handling of the above-cut-off atoms. Other common methods for many-body quantum systems, e.g. exact diagonalization approach~\cite{zhang_dong_2010} or multi-configuration time-dependent Hartree (MCTDH) method~\cite{beck_jackle_2000,alon_streltsov_2008},  are to date limited to investigations at zero-temperature or studies of equilibrium properties, whereas positive-P method~\cite{gilchrist_gardiner_1997,steel_olsen_1998} often has stability issues at long simulation times. It remains to be seen how important the violation of the Kohn mode actually is in practice in terms of other observables.

\section{\label{sec:thermalization}Example 2: Thermalization / Condensate Growth}
We next explore the interesting problem of thermalization, with direct relevance to the condensate growth from a partially-condensed initial state. For an out-of-equilibrium Bose gas, the interactions between the particles tend to redistribute energy among them. Except for the very few cases (e.g. the integrable one-dimensional systems~\cite{gring_kuhnert_2012,smith_gring_2013} or the Boltzmann monopole mode~\cite{boltzmann_1879,lobser_barentine_2015}), the non-equilibrium system is expected to rethermalize to a thermal equilibrium state, where populations of the energy modes are distributed according to the Bose-Einstein distribution. The evaporative cooling of a single-component Bose gas or the sympathetic cooling of a two-component Bose-Bose mixture are particularly well-suited examples to investigate the rethermalization dynamics.

\subsection{\label{sec:single_thermal}Single-component thermalization}
\begin{figure}
  \includegraphics[width=0.49\textwidth]{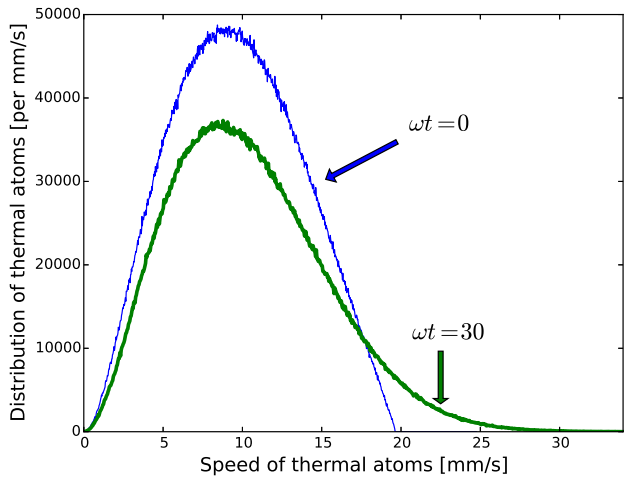}
  \caption{\label{fig:tphisto_final} Distributions of thermal atoms as a function of speed at the beginning (thin blue) and the end (thick green) of a simulation, after atoms with energy above the cutoff energy have been removed. $T_{\rm cut}=2400\,$nK. These distributions are sampled with $5.5\times10^6$ (begin) and $4.5\times10^6$ (end) test particles, respectively.}
\end{figure}

We first consider the evaporative cooling of a single-component Bose gas~\cite{miesner_stamper-kurn_1998}. At the final stage of the cooling process, a radio-frequency sweep is applied to quickly remove atoms with an energy above a certain cutoff energy. A non-equilibrium situation is thus created, where the atoms will interact and rethermalize in a completely-isolated environment, with a growth in the condensate number solely due to the internal dynamics of the Bose gas.

A ZNG-type simulation of the above scenario has been carried out by Bijlsma, Zaremba and Stoof~\cite{bijlsma_zaremba_stoof_2000} using the ergodic approximation for the thermal cloud and the Thomas-Fermi approximation for the condensate (see also the quantum kinetic treatment of Davis \etal~\cite{davis_gardiner_2000}). We report here results from full ZNG simulations without relying on these approximations. In particular, we show that (i) our numerical implementation can establish a thermal equilibrium state from a highly non-equilibrium situation, (ii) there is a smaller increase in the condensate number if the cutoff energy is too low and (iii) there is a two-stage dynamics that have not been discussed before in the past studies~\cite{miesner_stamper-kurn_1998,gardiner_lee_1998,davis_gardiner_2000,bijlsma_zaremba_stoof_2000}.

\begin{figure}
  \includegraphics[width=0.49\textwidth]{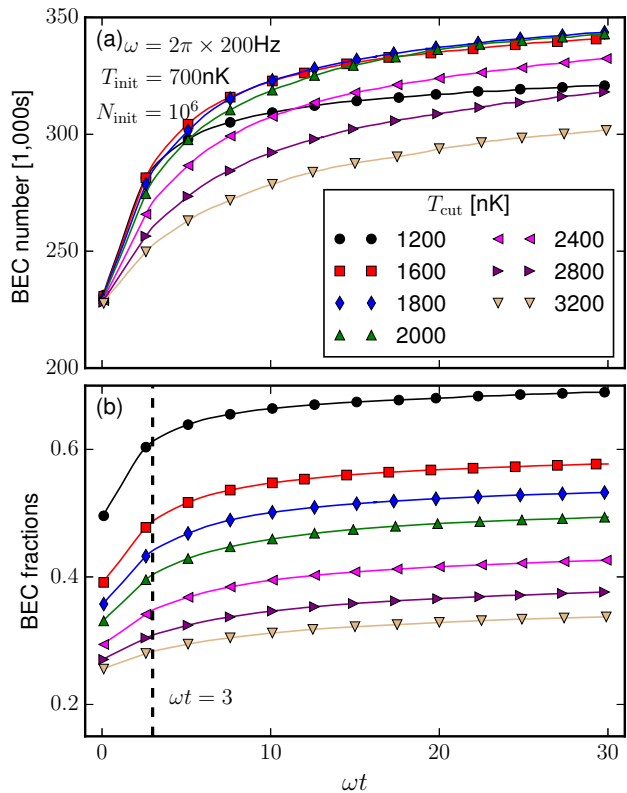}
  \caption{\label{fig:cutoff_fraction}Growth of (a) condensate number and (b) condensate fraction of $^{87}$Rb atoms in a harmonic trap with frequency $\omega=2\pi\times200\,$Hz as a function of time $t$ after thermal atoms with energy  greater than $k_B T_{\rm cut}$ are rapidly removed from the trap. Initially, there are a total of $10^6$ atoms at equilibrium with temperature $700\,$nK. The scattering length is $a_{\rm Rb}=99a_0$. The solid lines connect all data points, but only one out of twenty five data points are shown with symbols.}
\end{figure}

Our simulation begins with $10^6$ $^{87}$Rb atoms in an isotropic harmonic trap with angular trap frequency $\omega=2\pi\times200\,$Hz at a temperature of $700\,$nK (critical temperature $T^0_c=900\,$nK for a non-interacting gas). The scattering length is chosen to be $a_{\rm Rb} = 99a_0$. A self-consistent calculation with the Hartree-Fock approximation yields a condensate fraction of approximately 23\%. We first generate approximately $7.7\times10^6$ test particles, which are distributed according to the Bose-Einstein distribution, to simulate the thermal cloud. Subsequently, any test particle with energy $\varepsilon(\vec{r},\vec{v})$ [$=\frac{1}{2}m_{\rm Rb}\vec{v}^2 + U_n^{\rm Rb}(\vec{r})$, where $\vec{v}$ is the velocity of the test particle, $\vec{r}$ is the position of the test particle and $m_{\rm Rb}$ is the mass of a $^{87}$Rb atom] above a cutoff value $E_{\rm cut}=k_B T_{\rm cut}$ is removed from our simulation. This truncation process represents the rapid quench in experiments. It leaves approximately $6.7\times10^6$ test particles in the trap if $T_{\rm cut}=3200\,$nK, but only approximately $2.4\times10^6$ test particles if $T_{\rm cut}=1200\,$nK. \Eref{eq:dschro1} and \eref{eq:qbe} are then solved with this truncated test particles distribution as the initial thermal cloud distribution. An example of such an initial thermal cloud distribution as a function of the speed $|\vec{v}|$ for $T_{\rm cut}=2400\,$nK is shown as the thin blue line in \fref{fig:tphisto_final}. Instead of a sharp jump in the distribution (see e.g. figure 3 of~\cite{davis_gardiner_2000}), our distribution increases gradually once the speed falls below a cutoff value. This apparent difference stems from the fact that we are showing the truncated distribution as a function of the speed (hence integrating out the spatial dependence of energy $\varepsilon(\vec{r},\vec{v})$ on the effective potential $U_n^{\rm Rb}(\vec{r})$), while \cite{davis_gardiner_2000} showed the distribution as a function of the energy. For this same reason, while $T_{\rm cut}=2400\,$nK corresponds to a maximum speed of 21.4\,mm/s, the true maximum speed in the simulation is slightly below 20\,mm/s.

\begin{figure}
  \center
  \includegraphics[width=0.49\textwidth]{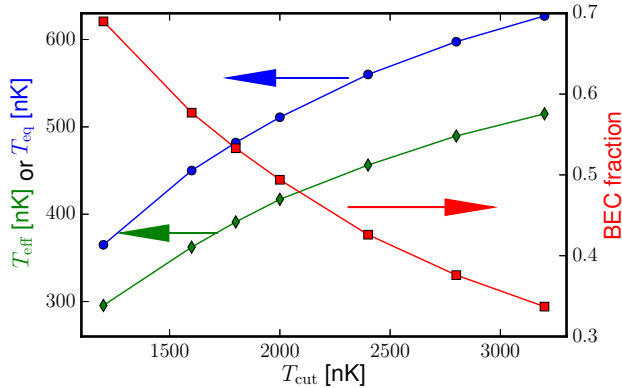}
  \caption{\label{fig:cutoff_finalT}Effective temperature $T_{\rm eff}$ of the thermal cloud (left axis, green diamonds), equilibrated temperature $T_{\rm eq}$ (left axis, blue circles) and BEC fractions (right axis, red squares) for different energy cutoff $k_B T_{\rm cut}$. Same parameters as \fref{fig:cutoff_fraction}. The effective temperature measures the width of the thermal cloud in velocity space. The equilibrated temperature is determined as the temperature that will give the same condensate number and total atom number from an equilibrium calculation as the dynamical simulation.}
\end{figure}

The removal of the high-energy particles decreases the total energy of the system. The thermal cloud then rethermalizes to produce a growth of the condensate number (\fref{fig:cutoff_fraction}) through the $C_{12}$ collisional process, and repopulates the high-energy modes (i.e. the exponentially-decreasing tail of the thick green line in \fref{fig:tphisto_final}) through the $C_{22}$ collisional process. It is interesting to note that, while the truncation process always leads to an increase in the condensate number, the magnitude of increase at the end of the simulation saturates at around $2000\,$nK as the cut becomes deeper. Eventually, this trend is reversed, and the condensate starts decreasing as the cut is deeper than $1800\,$nK (\fref{fig:cutoff_fraction}a), in qualitative agreement with experiments (see, e.g. \cite{anderson_ensher_1995}). Nevertheless, the condensate fraction, computed as the ratio of the condensate number over the total atom number at the end of simulation ($\omega t=30$), increases monotonically as we consider a deeper cut (\fref{fig:cutoff_fraction}b and \fref{fig:cutoff_finalT}), similar to our earlier findings with surface evaporative cooling \cite{markle_allen_2014}.

\begin{figure}
  \includegraphics[width=0.49\textwidth]{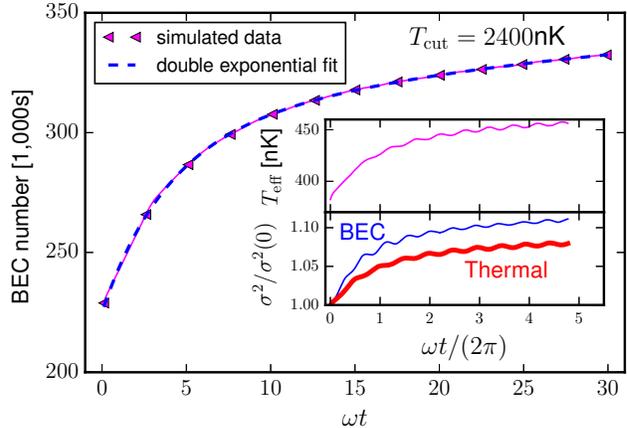}
  \caption{\label{fig:fit_Ec2400}Growth of the condensate number from a full simulation of the ZNG model (solid magenta) with an energy cutoff $k_B\times2400\,$nK. The simulation result is well-fitted by a double exponential~\eref{eq:double_exp}, shown as blue dashes. The two fitted time scales are $\omega\tau_1=4$ and $\omega\tau_2=27$. The parameters are the same as \fref{fig:cutoff_fraction}. Inset shows (top) the effective temperature of the thermal cloud~\eref{eq:T_eff} and (bottom) the normalized widths~\eref{eq:width} of the condensate (thin blue) and the thermal cloud (thick red). The solid line connects all data points, but only one out of twenty five data points are shown with symbols. See online supplementary video for the variation of density profiles with respect to time.}  
\end{figure}

In contrast to previous studies on condensate growth, which typically describe the rethermalization dynamics by a single thermalization time scale $\tau$ (with possibly also an onset time $\tau_{\rm onset}$ if the initial condensate fraction is negligible), our simulation data displays two thermalization time scales. This is revealed by a double exponential fit to the condensate number $N_c(t)$ as a function of time,
\begin{eqnarray}\label{eq:double_exp}
N_c(t) = N_i &+& (N_1-N_i)\times(1-\Exp{-t/\tau_1}) \\\nonumber
                    &+& (N_{\rm eq}-N_1)\times(1-\Exp{-t/\tau_2}).
\end{eqnarray}
The first time scale, $\tau_1$, is associated with the rapid growth in the condensate number. For our current parameters, this is approximately given by $\omega\tau_1\approx3$ (vertical dashes in \fref{fig:cutoff_fraction}b). The second time scale $\tau_2$ is an order of magnitude larger and its origin is not yet fully resolved. Similar numerical condensate growth curves that could possibly be fitted with the double exponential form~\eref{eq:double_exp} are also found in earlier works that employ the classical field approach~\cite{proukakis_schmiedmayer_2006} or the ZNG approach~\cite{bijlsma_zaremba_stoof_2000,jackson_zaremba_2002a}, even though the double exponential fit is not explicitly mentioned in any of those studies. In \fref{fig:fit_Ec2400}, we show the excellent agreement between the double exponential fit (blue dashes) and the simulated data (magenta solid line) for $T_{\rm cut}=2400\,$nK, with fitted time scales $\omega\tau_1=4$ and $\omega\tau_2=27$. Similar agreement between the numerical data and the fitted curves are also obtained for the other values of $T_{\rm cut}$. In \ref{sec:compare_fit}, we further compare the residuals using different fitting functions to convincingly demonstrate that the double exponential fit~\eref{eq:double_exp} is indeed the best choice.

We note that the two-stage dynamics reported here is quite different from the various two-stage dynamics that have been studied before. Most of the previous works deal with the early stage of condensate growth where the condensate is negligible. For examples, Kagan \etal~\cite{kagan_svistunov_1992,svistunov_2001} consider the formation of a quasi-condensate with phase-fluctuations that die out to produce the truly phase-coherent condensate; Bose-stimulated process is considered to arrive at a relaxation process with an onset time~\cite{miesner_stamper-kurn_1998}; the two-stage dynamics observed in~\cite{kohl_davis_2002} is attributed to the slow ergodic mixing at the very early stage of condensate growth; or a two-stage condensation dynamics due to the different transverse and axial trapping energies~\cite{van_druten_ketterle_1997,gorlitz_vogels_2001}.

In comparison, our simulations start off with an appreciable condensate fraction ($23\%$) and is accompanied by the emergence of a monopole oscillation at the later stage. This appearance of the monopole oscillation is reminiscent of the quadrupole mode observed in the growth of an elongated condensate~\cite{shvarchuck_buggle_2002,hugbart_retter_2007,kuwamoto_hirano_2012}. 
The top panel of the inset of \fref{fig:fit_Ec2400} shows the effective temperature $T_{\rm eff}$ of the thermal cloud, 
\begin{equation}\label{eq:T_eff}
  T_{\rm eff} = \frac{2}{3k_B}\times\frac{1}{N_{\rm tp}}\sum^{N_{\rm tp}}_{i} \frac{\vec{p}_i^2}{2m},
\end{equation}
as a function of the simulation time $t$, where $T_{\rm eff}$ measures the average kinetic energy of the $N_{\rm tp}$ test particles. The bottom panel of the inset shows the widths of the condensate and the thermal cloud normalized to the initial values, defined as
\begin{equation}\label{eq:width}
  \sigma^2(t) = \int d\vec{r}\,\vec{r}^2\,n(\vec{r},t),
\end{equation}
and $n(\vec{r},t)=n_c(\vec{r},t)$ or $\tilde{n}(\vec{r},t)$. The oscillatory structure with an angular frequency of approximately $2\omega$ appearing after $\tau_1$ hints at the presence of a monopole mode. Further evidence of the monopole oscillation is given in~\ref{sec:extract_mode}, where we extract the monopole and quadrupole oscillations from a double exponential fit of the condensate widths as well as the thermal cloud widths, but the extracted data are not able to provide conclusive evidence with regards to the potential relationship between the second time scale $\tau_2$ and the monopole oscillation. In view of the recent results on the Boltzmann monopole mode~\cite{boltzmann_1879,lobser_barentine_2015,straastma_colussi_2016}, whether this two-stage dynamics remains valid for an anisotropic trap would be an interesting subject for future studies. 

At this point, it is worth noting that the increase in the effective temperature $T_{\rm eff}$ of the thermal cloud (inset of \fref{fig:fit_Ec2400}) might not be apparent at first sight. This is because we typically associate cooling with a decrease in the temperature. The key idea is that it only makes sense to use $T_{\rm eff}$, which measures the width of the thermal cloud in the velocity space, to characterize the degree of coldness when the system reaches equilibrium. In fact, this increase of $T_{\rm eff}$ in time occurs naturally as the speed distribution of the thermal atoms relaxes from a truncated distribution (thin blue line in \fref{fig:tphisto_final}), which has a smaller width because of the lack of high-energy population, to an equilibrium distribution (thick green line in \fref{fig:tphisto_final}), which has an exponential tail that extends far into the high-energy region. A plot of $T_{\rm eff}$ at the end of our simulation versus the cutoff energy (green diamonds in \fref{fig:cutoff_finalT}) shows that a deeper cut consistently yields a cooler equilibrium system.

On the other hand, there is also an interesting increase in the spatial width of the thermal cloud (\fref{fig:fit_Ec2400}). This is because thermal atoms further away from the trap center are more likely to be removed by the truncation process, hence a pressure difference is created to push the thermal cloud outwards once the dynamical simulation commences. The outward expansion of the thermal cloud in turn can drive the monopole mode oscillation of the condensate through the mean field potential. An immediate consequence of this picture is that, the deeper is the cut (i.e. smaller $T_{\rm cut}$), the larger is the pressure difference, hence the larger is the oscillation amplitude (see also \ref{sec:extract_mode}).

\begin{figure}
  \includegraphics[width=0.49\textwidth]{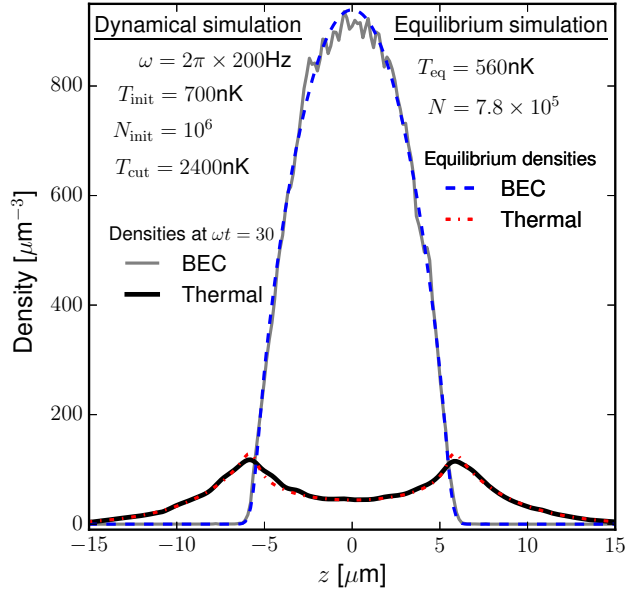}
  \caption{\label{fig:cutoff_compareDensity}Density profiles of condensate (thin grey) and thermal cloud (thick black) at the end of the thermalization simulation ($\omega t=30$) with $3.3\times10^5$ condensate atoms. The small wiggles and the slight asymmetry in the density profiles are statistical fluctuations that originate from the Monte Carlo sampling of both the initial thermal cloud distribution and the collisional integrals. Same parameters as \fref{fig:cutoff_fraction} with an energy cutoff of $k_B\times2400\,$nK, which leave a total of $7.8\times10^5$ $^{87}$Rb atoms in the trap. The blue dashes (red dash-dots) give the equilibrium condensate (thermal cloud) density at temperature $560\,$nK with the same total number of atoms, where the temperature is chosen such that the condensate number is also the same as the dynamical simulation.} 
\end{figure}

In spite of the long relaxation time $\tau_2$, we find that it is still possible to assign an equilibrium temperature $T_{\rm eq}$ to the system at the end of the equilibration process ($\omega t=30$)~\footnote{We expect the system to reach its true equilibrium after about $10\tau_2$. However, numerically simulating up to this time would require a computational time of one to two months. Since the condensate number is expected to increase by at most ten percent if we perform simulation from $\omega t=30$ to $\omega t=300$, we believe that our simulations running up to $\omega t=30$ is sufficient for the present purpose.}. This is done by searching for a temperature that yields the same condensate number and total atom number in our equilibrium calculations. The condensate density and the thermal cloud density obtained from our equilibrium calculation (dashes and dash-dots) are plotted against the density profiles at $\omega t=30$ of the dynamical simulation (solid lines) in \fref{fig:cutoff_compareDensity} and show remarkable agreement. The decrease in $T_{\rm eq}$ for deeper cut (\fref{fig:cutoff_finalT}) is consistent with our intuition and reaffirms the validity of the ZNG model.

\subsection{Two-component thermalization}
We now consider the sympathetic cooling of a two-component mixture~\cite{myatt_burt_1997}. In a typical experimental situation, the first component can be easily cooled, e.g. via the application of a radio-frequency sweep, while the second component is cooled by being in thermal contact with the first component. Atoms of different components can therefore collide elastically and exchange energy as long as the two components overlap in space. The thermalization rate of such a scenario has been investigated both experimentally~\cite{delannoy_murdoch_2001} and theoretically~\cite{lewenstein_cirac_1995,timmermans_cote_1998,geist_you_1999,delannoy_murdoch_2001,papenbrock_salgueiro_2002}. However, these theoretical estimates often rely on approximations to simplify the calculations, such as 
\begin{enumerate}
\item the high-temperature approximation where all atoms follow a Maxwell-Boltzmann distribution, or 
\item the omission of mean-field contribution, where the condensate wavefunction is identified as the ground state of a harmonic oscillator. 
\end{enumerate}
A thorough investigation that takes into account the presence of \emph{both} condensates self-consistently is still missing.

We aim to eventually fill this gap by performing systematic studies of sympathetic cooling using our two-component ZNG model. Due to the many different collisional processes involved (eight collisional processes in a two-component mixture \cite{edmonds_lee_2015a,edmonds_lee_2015b} versus two collisional processes in a single-component Bose gas), the dynamics is much more complicated and also much more interesting to study.

\begin{figure}
  \includegraphics[width=0.49\textwidth]{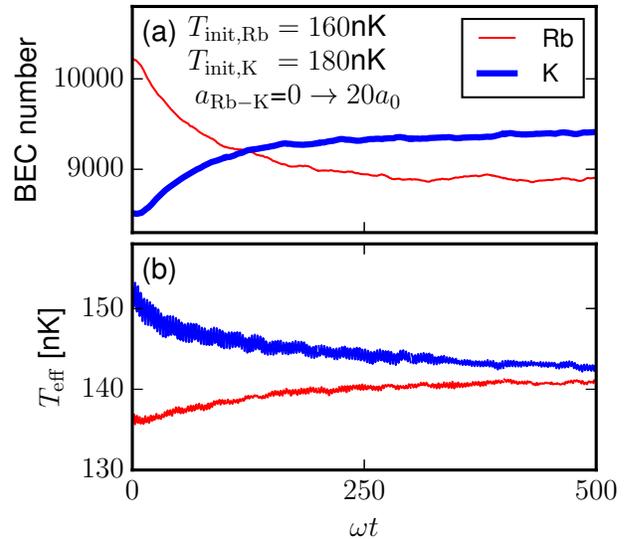}
  \caption{\label{fig:scaled_colision} Sympathetic cooling of a $^{87}$Rb-$^{41}$K mixture with $2\times10^4$ atoms each in isotropic trap with frequency $\omega=2\pi\times200\,$Hz. The two components have initial temperatures 160$\,$nK ($^{87}$Rb) and 180$\,$nK ($^{41}$K), with intra-component scattering length $a_{\rm Rb}=99a_0$ and $a_{\rm K}=60a_0$. The inter-component scattering length $a_{\rm Rb-K}$ is linearly increased from 0 to $20a_0$ in 10$\,$ms to initiate the cooling process. The condensate-exchange collisional integral $\mathds{C}_{12}^{kj}$ is excluded due to its much shorter time scale compared to other collisional processes. The inter-component thermal-thermal collisions $C_{22}^{kj}$ has been scaled up by 50 times while the inter-component thermal-condensate collisions $C_{12}^{kj}$ has been scaled up by 10 times. Figure shows (a) condensate numbers and (b) effective temperatures~\eref{eq:T_eff} as $^{87}$Rb atoms sympathetically cool $^{41}$K atoms.} 
\end{figure}

\begin{figure}
  \includegraphics[width=0.49\textwidth]{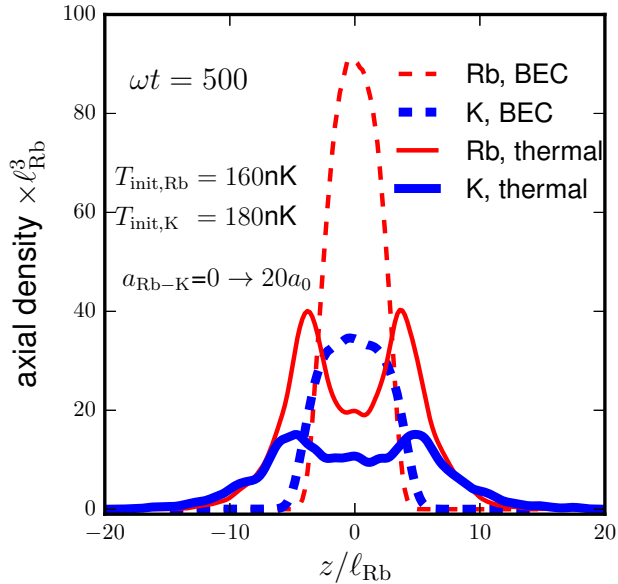}
  \caption{\label{fig:2compo_final_density}Density profiles along the axial axis for $^{87}$Rb atoms (thin red lines) and $^{41}$K atoms (thick blue lines) at the end of the cooling process ($\omega t=500$), where dashes refer to the condensates and solid lines refer to the thermal clouds. The thermal clouds have been magnified by 10 times to increase clarity. The small wiggles and the slight asymmetry in the density profiles are statistical fluctuations that originate from the Monte Carlo sampling of both the initial thermal cloud distribution and the collisional integrals. Same parameters as \fref{fig:scaled_colision}.} 
\end{figure}

In \fref{fig:scaled_colision}, we show that our two-component ZNG model can establish an equilibrium solution in a sympathetic cooling setup. We start with a $^{87}$Rb-$^{41}$K mixture, each with 20,000 atoms in an isotropic harmonic trap with identical trap frequency $\omega=2\pi\times200\,$Hz. Initially, there is no thermal contact between the two components (i.e. inter-component scattering length $a_{\rm Rb-K}=0$). The intra-component scattering lengths are $a_{\rm Rb}=99a_0$ and $a_{\rm K}=60a_0$. The two components have different initial temperatures, $T_{\rm Rb}=160\,$nK and $T_{\rm K}=180\,$nK, and different condensate fractions (51\% for $^{87}$Rb and 43\% for $^{41}$K). The critical temperatures are $T_{c,{\rm Rb}}=226\,$nK and $T_{c,{\rm K}}=233\,$nK when taking into account of the finite-size corrections and mean-field corrections~\cite{dalfovo_giorgini_1999}.

In order to initiate the cooling process, we linearly ramp up the inter-component scattering length $a_{\rm Rb-K}$ from 0 to $20a_0$ in 10$\,$ms, and maintain the value afterwards. In this case, the mixture remains miscible [$g_{\rm Rb-K}^2/(g_{\rm Rb} g_{\rm K})=0.3$] such that there is a large spatial overlap of the two components (see \fref{fig:2compo_final_density} for the density profiles at $t=500/\omega\approx400\,$ms). For realistic values of scattering lengths, the thermalization time scales are of the order of seconds, which translate into a very long computational time~\footnote{A typical simulation up to $\omega t = 500$ takes 10 days on a 20-core machine, with Intel Xeon CPU@3GHz.}. In order to speed up the equilibration process within our simulations, we increases the inter-component thermal-thermal collisional cross-section by 50 times, and the inter-component thermal-condensate collisional cross-section by 10 times when sampling the collision events. In practice, this means that we make the substitutions
\begin{equation}
R_{12}^{kj}\to \kappa_{12} R_{12}^{kj}
\end{equation}
in equation~\eref{eq:dschro1} and
\begin{equation}\eqalign
C_{22}^{kj}\to \kappa_{22} C_{22}^{kj},\\
C_{12}^{kj}\to \kappa_{12} C_{12}^{kj}
\end{equation}
in equation~\eref{eq:qbe} for $k\neq j$, $\kappa_{22}=50$ and $\kappa_{12}=10$. In addition, we omit the condensate-exchange process ($\mathds{C}_{12}^{kj}$ and $\mathds{R}_{12}^{kj}$) in our simulation as its time scale is typically an order of magnitude shorter than the other collisional processes~\cite{edmonds_lee_2015a,edmonds_lee_2015b}.

\Fref{fig:scaled_colision}(a) shows that the condensate number of $^{87}$Rb atoms decreases in time and this is accompanied by a growth in the condensate number of $^{41}$K atoms, as $^{87}$Rb atoms sympathetically cool $^{41}$K atoms. \Fref{fig:scaled_colision}(b) shows the corresponding change in the effective temperatures of the thermal clouds~\eref{eq:T_eff}. The fact that the condensate numbers saturate to finite values and the two effective temperatures converge shows that an equilibrium situation is established at the end of our simulation ($\omega t=500$).

Both condensate numbers in \fref{fig:scaled_colision}(a) are well fitted by an exponentially-decaying function,
\begin{equation}
N^j_c(t) = N_i^j + (N_f^j-N_i^j)\times(1-\Exp{-t/\tau_j}),
\end{equation}
with the thermalization time scales $\omega\tau_{\rm Rb}=83$ and $\omega\tau_{\rm K}=76$. These time scales are to be compared with the prediction based on Maxwell-Boltzmann distribution~\cite{delannoy_murdoch_2001},
\begin{equation}\label{eq:2compo_predict}
\frac{1}{\tau} = \frac{N_{\rm Rb}+N_{\rm K}}{3k_B (T_{\rm Rb} + T_{\rm K})/2}\frac{\omega^3 \sigma_{\rm Rb-K} \mathcal{M}}{2\pi^2}\times\kappa_{22},
\end{equation}
where $\sigma_{\rm Rb-K}=4\pi a_{\rm Rb-K}^2$ is the cross-section, the factor $\kappa_{22}$ takes into account that we have scaled up our collisional probabilities to speed up the computation, and $\mathcal{M} = [8(m_{\rm Rb}+m_{\rm K})^2]/[(m_{\rm Rb}+m_{\rm K})^3]$. This yields the estimate $\omega\tau\approx40$ that is comparable to the simulated time scales. This agreement could stem from the fact that we have scaled up the inter-component thermal-thermal collisional probabilities, making all collisional processes to have comparable time scales within our simulations. 

\begin{figure}
  \includegraphics[width=0.49\textwidth]{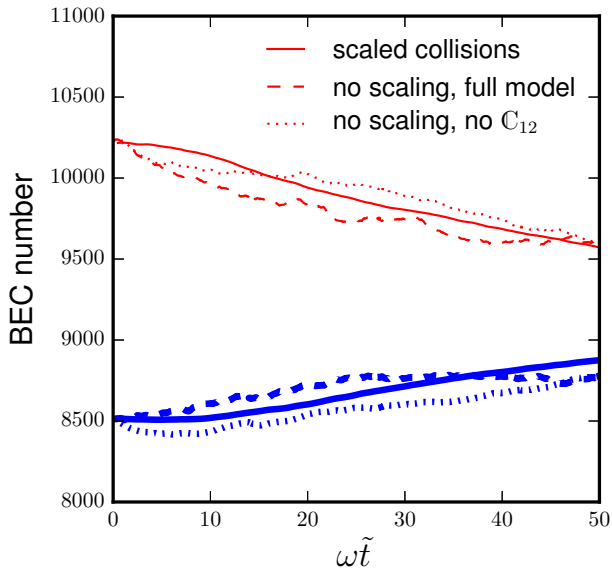}
  \caption{\label{fig:2compo_compare_collision}Variation of condensate numbers as a function of rescaled time $\tilde{t}$ instead of the true time $t$, as $^{87}$Rb atoms (red) sympathetically cool $^{41}$K atoms (blue). Same parameters as \fref{fig:scaled_colision}. Solid line shows simulation results with scaled collisional probabilities ($\kappa_{22}=50$, $\kappa_{12}=10$), omitting the condensate-exchange collisions, and $\tilde{t}=t$. Both dashed and dotted lines show results with true collisional probabilities ($\kappa_{22}=1$, $\kappa_{12}=1$) with $\tilde{t}=t/10$, where one include (dashed) while the other omit (dotted) the condensate-exchange process.} 
\end{figure}

It is interesting to see if such an agreement remains when we restore the true collisional probabilities. In \fref{fig:2compo_compare_collision}, we compare simulations with scaled probabilities (solid lines, $\kappa_{22}=50$, $\kappa_{12}=10$) and those with the true collisional probablities ($\kappa_{22}=1$, $\kappa_{12}=1$) that includes (dashed lines) or omits the condensate-exchange $\mathds{C}_{12}$ collisions (dotted lines). Our numerical results reveal that the condensate-exchange collision leads to faster thermalization at the initial stage, but it has less impact at the later stage, where the condensate numbers vary at comparable time scales in the presence or absence of the condensate-exchange collisions. In order to deduce the true thermalization time, the condensate numbers are plotted in terms of the rescaled time $\tilde{t}$, where $\tilde{t}=t$ for the simulation with scaled probabilities, but $\tilde{t}=t/10$ for the simulations with true probabilities. The similarity of the curves implies that the true thermalization time scale is likely to be of the order of $\omega\tau=800$, shorter than the $\omega\tau=2000$ predicted by equation~\eref{eq:2compo_predict}.

As we use realistic numbers in our simulation, we expect the two-component thermalization time in experiments to be of the order $800/\omega\approx 0.6s$, which is in fact comparable to typical waiting times in experiments. Taking into account that the overlap of two components is reduced due to the relative shift of trap minima, our simulation results therefore suggest that the two components may not necessarily have fully thermalized to a common temperature in experiments. This interesting question clearly merits further detailed investigation, both from a theoretical and an experimental perspective.

An important factor to consider here is that on one hand one seeks a large inter-component interaction strength $g_{12}$ in order to speed up the $C_{12}^{kj}$ ($k\neq j$) collisions that dominate the process of sympathetic cooling; however, a larger $g_{12}(>0)$ leads to enhanced phase separation, thus minimizing the effective overlap over which thermalization / cooling takes place, resulting in a slower sympathetic cooling rate. A critical balance between those two competing mechanisms is here needed to optimize the cross-thermalization and the sympathetic cooling process efficiently.

\section{Outlook}
We have demonstrated that our numerical implementation of the ZNG model can capture the essential physics of non-equilibrium Bose-Einstein condensates at finite temperature, namely the Kohn mode and the rethermalization dynamics associated with condensate growth for both single and binary atomic gases. 

Revisiting the well-studied single-component problem, but for the specific case of a spherically-symmetric (isotropic) trap, we have observed two interesting features which deserve further attention, also from an experimental point of view. Firstly, we found that, following a rapid evaporative cooling quench, the system grows to a state with higher condensate fraction on two distinct timescales, a rapid one associated with rapid condensate number growth, and a much slower one. Analogous, yet physically distinct, two-stage condensate formation dynamics have been previously reported in various different contexts, e.g. associated with the excitation of a quadrupolar mode for highly-elongated systems. Parallel to this, we have found that the condensate growth process in an isotropic trap naturally excites the monopole mode, which in this geometry is long-lived and so could be experimentally observable (whereas in our case, excitation of the quadrupolar mode is suppressed). It would thus be very interesting to see an experimental study of controlled condensate growth, based on rapid evaporative cooling, as a function of trap aspect ratio. Extrapolating beyond our findings, one could envisage a situation whereby the condensation process leads either to a quadrupolar mode excitation for very elongated traps (for which monopole excitation is suppressed), or to a monopolar excitation for isotropic traps, with intermediate geometries having a variable amount of excitation and decay timescales for the different excitation modes. It would also be interesting to study whether the excited modes are in-phase or out-of-phase for the condensate and thermal cloud, a problem significantly more complicated than the previously conducted controlled excitation experiments~\cite{jin_ensher_1996,lobser_barentine_2015,straastma_colussi_2016}, as the condensate fraction and number are constantly changing during the condensate growth. The possiblity of exciting different modes following shock-cooling based on system geometry thus appears to be well-worth analyzing further.

Interesting open problems can also be explored using the two-component ZNG model, with the model predictions benchmarked against experimental results. Two particularly attractive research directions concern the controlled studies of the collective modes of a mixture and the sympathetic cooling in the presence of a partially-condensed Bose gas.

The former has gathered increasing experimental interest~\cite{ferrier-barbut_delehaye_2014,eto_takahashi_2015a,li_zhu_2015,bienaime_fava_2016} recently, but a major experimental challenge remains in reducing the relative shift of the trapping-potential minima. This relative shift can arise if the two components experience different spring constants of the harmonic potentials, or if the two components have different masses~\cite{wacker_jorgensen_2015}. Any experimental achievement in minimizing/eliminating the trap sag would immediately open up the possibility to study the effect of inter-component interaction on collective modes, such as the number-dependent miscible-immiscible transition~\cite{lee_jorgensen_2016} or the bifurcation of single-component collective modes~\cite{kasamatsu_tsubota_2004b}. In addition, the presence of the thermal clouds, which could possibly reach the hydrodynamic regime, leads to a `four-fluid' model of the binary mixture with more complicated dynamics, a glimpse of which can be found in \cite{edmonds_lee_2015a,edmonds_lee_2015b}.

On the other hand, while sympathetic cooling has become a routine procedure in cold-atom experiments, the theoretical investigation of the sympathetic-cooling rates below the critical temperature is still quite limited. Interesting questions that revolve around the thermalization time scales include the role of the condensates~\cite{timmermans_cote_1998} and the effect of the spatial overlap between the two components (which is in turn affected by the trap sag or the miscibility). The eight collisional processes with disparate collisional time scales~\cite{edmonds_lee_2015a,edmonds_lee_2015b} also lead to the fascinating possibility of quasi-equilibrium dynamics~\cite{nikuni_zaremba_1999}. How the different collisional processes can enhance or suppress each other, and how this problem can be explored experimentally, are open questions to be answered.

Finally, we would like to highlight that our atomistic simulation of the two-component ZNG model are not limited to study a Bose-Bose mixture. With the appropriate modifications to the quantum Boltzmann equations~\eref{eq:qbe} and the mean-field potentials~\eref{eq:effUc} and \eref{eq:effUt}, we can also study the non-equilibrium dynamics of a Bose-Fermi mixture, including the case where both the bosons and the fermions exist in the superfluid phase~\cite{ferrier-barbut_delehaye_2014}, or a partially-condensed Bose gas immersed in a spin-polarized Fermi sea~\cite{truscott_strecker_2001,schreck_khaykovich_2001}.

\ack
KLL and NPP thank E. Zaremba and C. J. E.  Straatsma for their critical reading of the manuscript, and acknowledge discussion with M. J. Edmonds, A. J. Allen, S. A. Gardiner and T. P. Billam, and support from EPSRC Grant No. EP/K03250X/1. This work made use of the facilities of N8 HPC Centre of Excellence, provided and funded by the N8 consortium and EPSRC (Grant No.EP/K000225/1). The Centre is co-ordinated by the Universities of Leeds and Manchester. We gratefully acknowledge the support of NVIDIA Corporation with the donation of the Tesla K40 GPU used for this research. Data supporting this publication is openly available under an `Open Data Commons Open Database License'. Additional metadata are available at: http://dx.doi.org/10.17634/122626-4. Please contact Newcastle Research Data Service at rdm@ncl.ac.uk for access instructions.

\appendix

\section{\label{sec:compare_fit}Comparison of different fitting functions to the condensate growth curve}
\begin{figure}[!h]
  \includegraphics[width=0.49\textwidth]{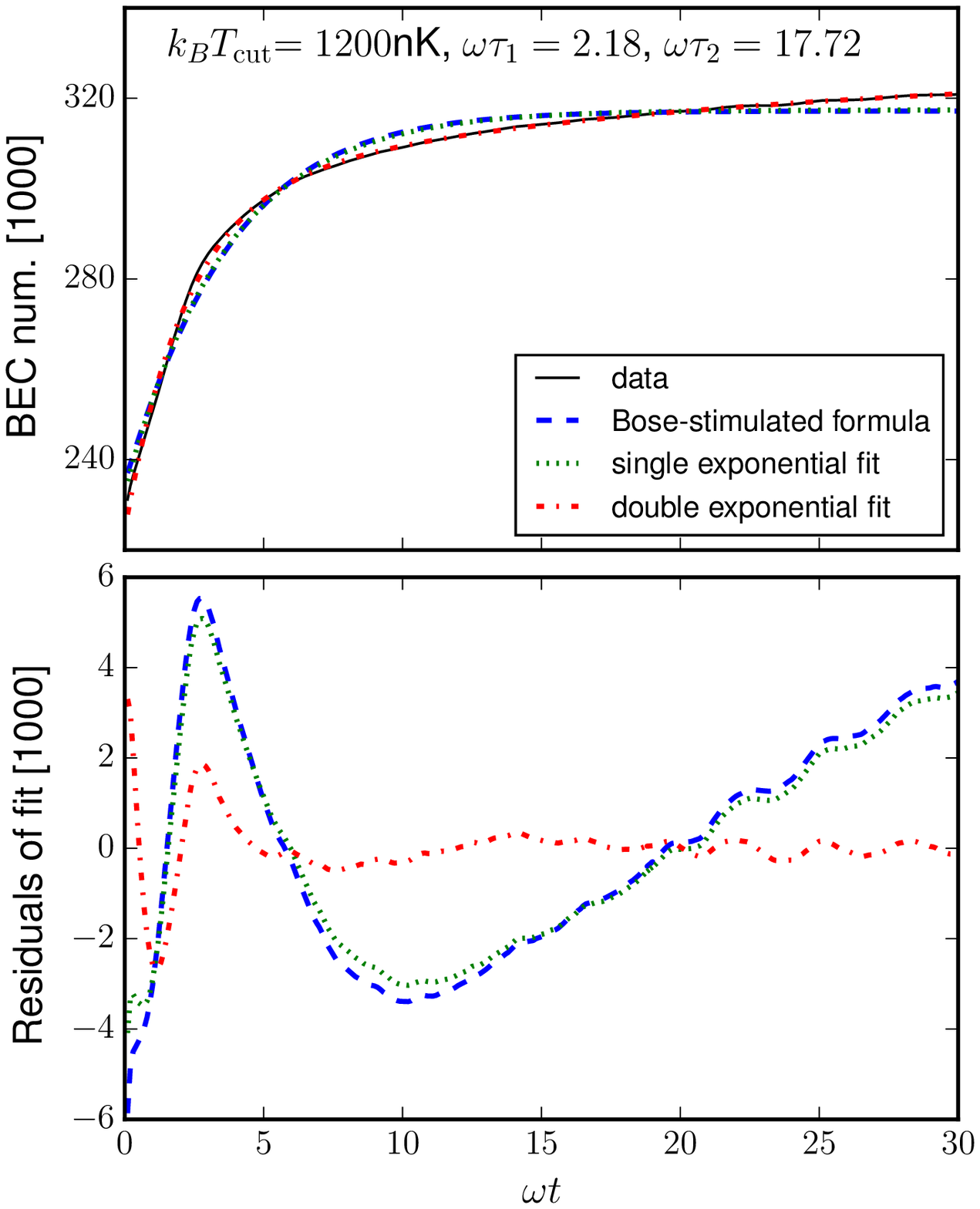}
  \caption{\label{fig:fitNc_cutoff1200}(top) Simulated condensate growth (black solid line) and the fitted curves using the Bose-stimulated formula~\eref{eq:bose_stimu} (blue dashed line), single exponential fit~\eref{eq:single_exp} (green dotted line) or double exponential fit~\eref{eq:appendix_double_exp} (red dash-dots) for cutoff energy $k_B T_{\rm cut}=1200\,$nK. Same parameters as \fref{fig:cutoff_fraction}. (bottom) Residuals~\eref{eq:residual} of the various fitting functions.} 
\end{figure}

\begin{figure}[!h]
  \includegraphics[width=0.49\textwidth]{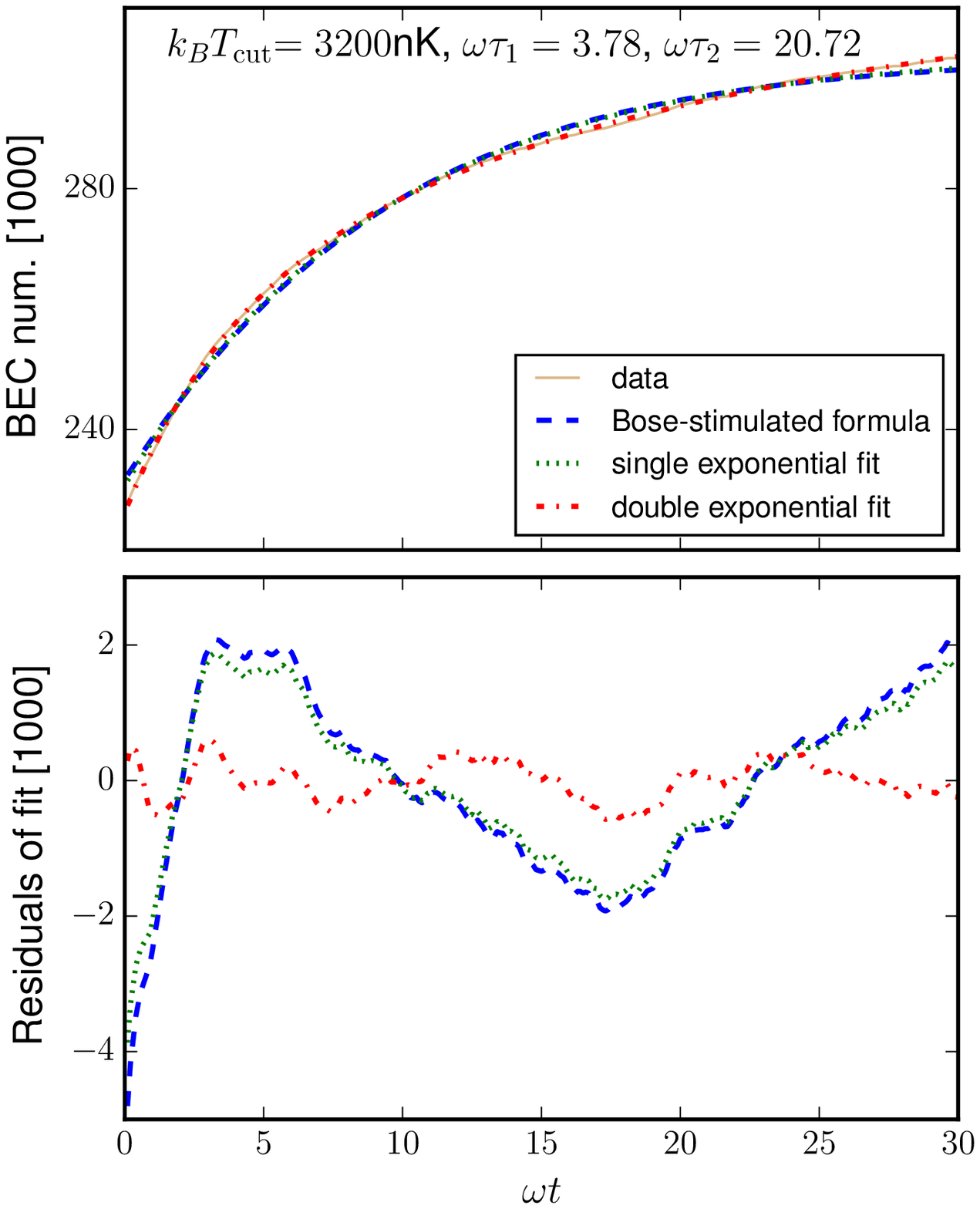}
  \caption{\label{fig:fitNc_cutoff3200}Same as \fref{fig:fitNc_cutoff1200} but for cutoff energy $k_B T_{\rm cut}=3200\,$nK.} 
\end{figure}
The condensate growth curves of a single-component Bose gas are fitted with three different functions:
\begin{enumerate}
\item A two-stage formula due to Bose-stimulated process~\cite{miesner_stamper-kurn_1998}, allowing for an initiation time,
\begin{equation}\label{eq:bose_stimu}
N_c(t) = N_i\Exp{t/\tau}[1+(N_i/N_f)^\delta(\Exp{\delta t/\tau}-1)]^{-1/\delta}
\end{equation}
for $\delta=2/5$,
\item a single exponential function that is commonly used to study relaxation processes,
\begin{equation}\label{eq:single_exp}
N_c(t) = N_i + (N_f-N_i)\times(1-\Exp{-t/\tau}),
\end{equation}
\item a double exponential function presumably considered but never explicitly investigated,
\begin{eqnarray}\label{eq:appendix_double_exp}
N_c(t) = N_i &+& (N_1-N_i)\times(1-\Exp{-t/\tau_1}) \\\nonumber
                    &+& (N_{\rm eq}-N_1)\times(1-\Exp{-t/\tau_2}).
\end{eqnarray}
\end{enumerate}

The fitted functions are shown in the top panels of \fref{fig:fitNc_cutoff1200} and \fref{fig:fitNc_cutoff3200} for two different cutoff energies. In order to assess the goodness of fit, we plot the residuals of fit in the bottom panels, defined as 
\begin{equation}\label{eq:residual}
\textrm{residual} = \textrm{data} - \textrm{fitted value}.
\end{equation}
Since the residuals of the double exponential functions (red dash-dots) always remain small compared to the other two functions, we conclude that our simulated data is indeed well described by a double exponential function.

\section{\label{sec:extract_mode}Extraction of monopole and quadrupole modes}
\begin{figure}
  \includegraphics[width=0.49\textwidth]{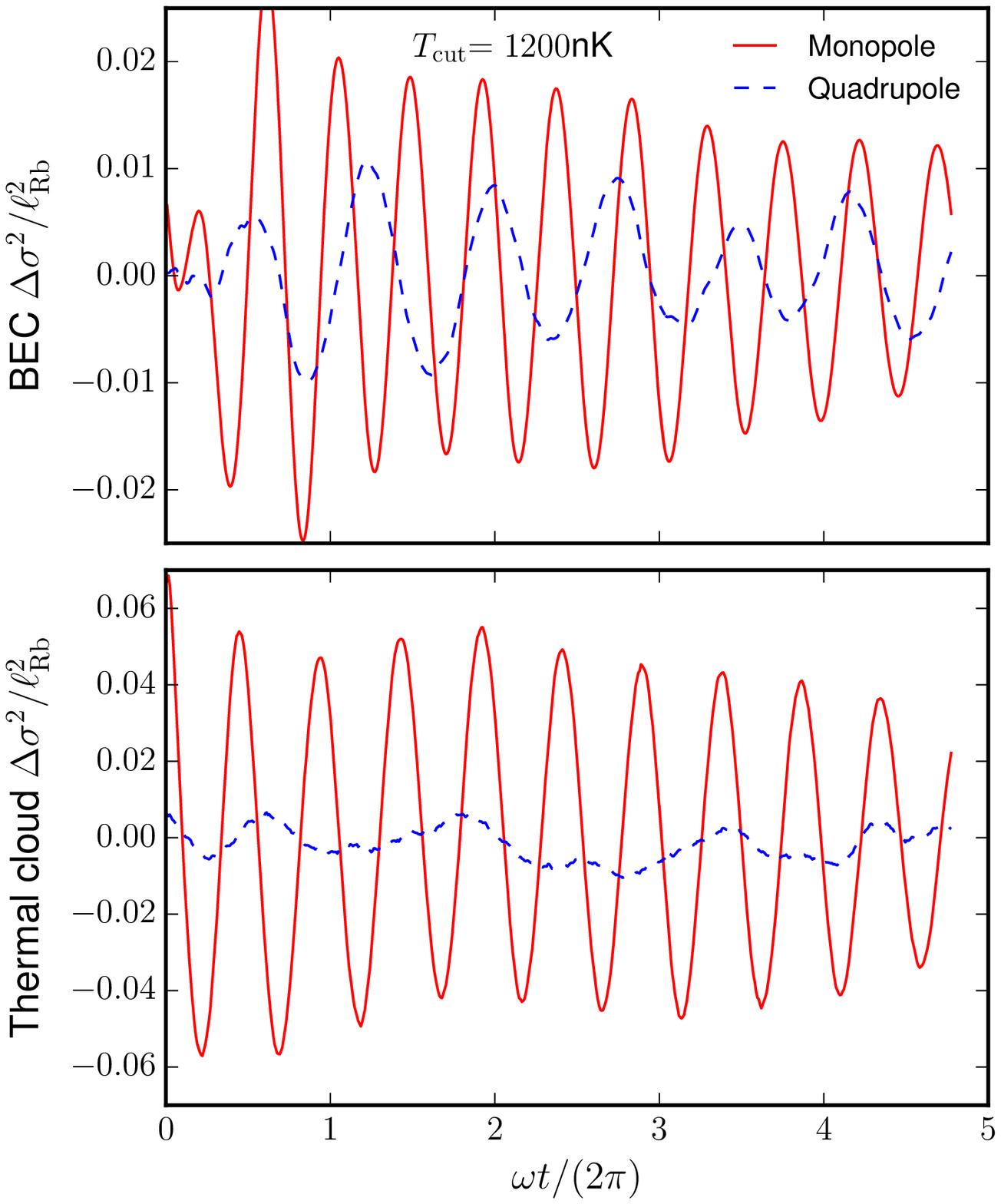}
  \caption{\label{fig:mode_Ec1200}Width fluctuations that measure monopole (solid red lines)~\eref{eq:mono} and quadrupole (dashed blue lines)~\eref{eq:quad} oscillations of (top) condesate and (bottom) thermal cloud for cutoff energy $k_B T_{\rm cut}=1200\,$nK. Same parameters as \fref{fig:cutoff_fraction}. } 
\end{figure}

\begin{figure}
  \includegraphics[width=0.49\textwidth]{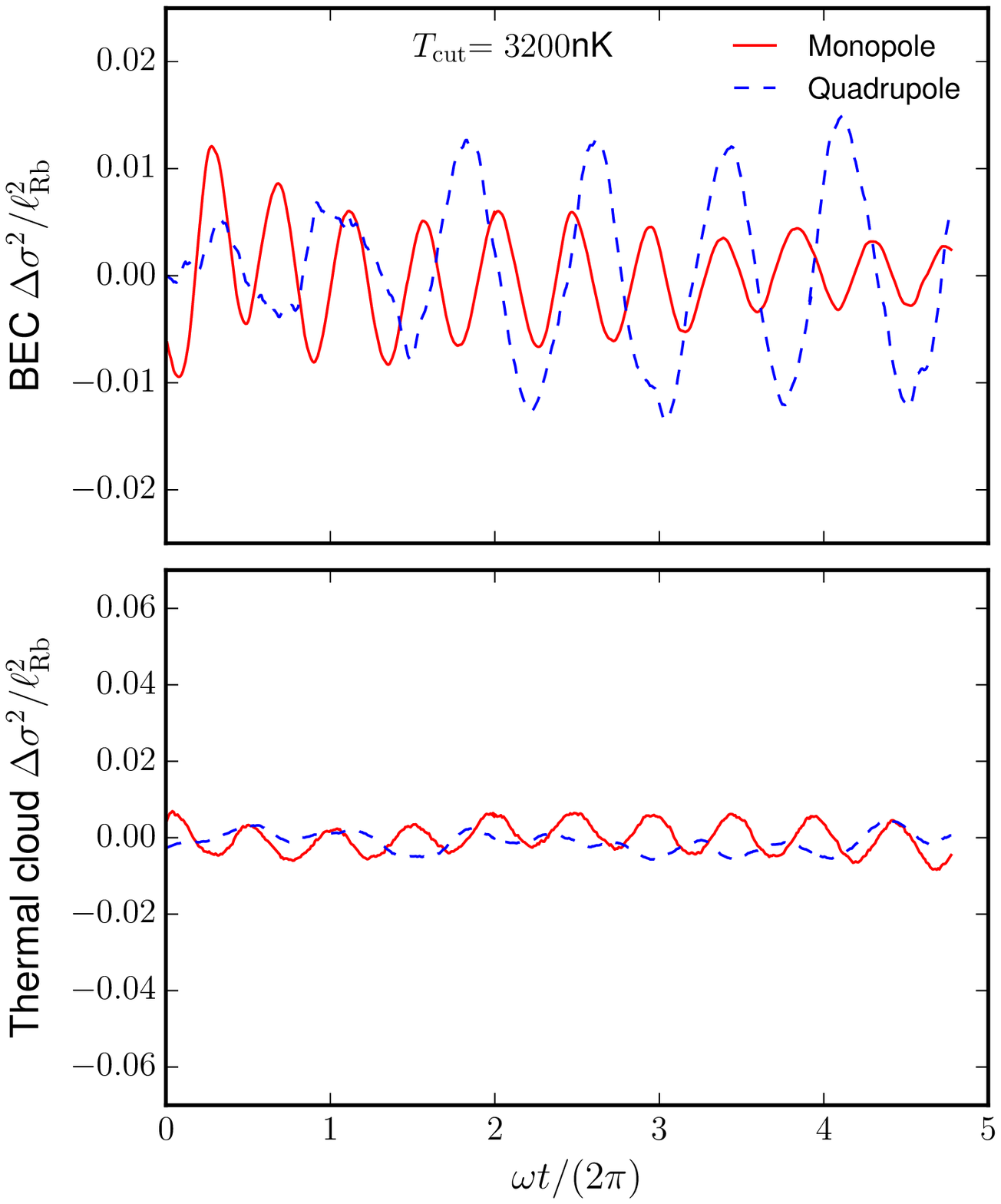}
  \caption{\label{fig:mode_Ec3200} Same as \fref{fig:mode_Ec3200} for cutoff energy $k_B T_{\rm cut}=3200\,$nK.} 
\end{figure}

We extract the monopole and quadrupole oscillation from the widths of the condensate as well as the thermal clouds. The background $\sigma^2_{\rm bg}(t)$ is first determined by fitting the radial width $\sigma^2(t)/3$ \eref{eq:width} by the double exponential function~\eref{eq:appendix_double_exp}, but with the time scales $\tau_1$ and $\tau_2$ as input parameters, where their values are determined by the fit of condensate numbers.

Width along each direction is calculated as 
\begin{numparts}
\begin{eqnarray}\label{eq:width_dir}
  \sigma_x^2(t) = \int d\vec{r}\,x^2\,n(\vec{r},t),\\
  \sigma_y^2(t) = \int d\vec{r}\,y^2\,n(\vec{r},t),\\
  \sigma_z^2(t) = \int d\vec{r}\,z^2\,n(\vec{r},t),
\end{eqnarray}
\end{numparts}
where $n(\vec{r},t)=n_c(\vec{r},t)$ or $\tilde{n}(\vec{r},t)$, with the corresponding fluctuations calculated as 
\begin{equation}
\Delta \sigma_i^2(t) = \sigma_x^2(t) - \sigma^2_{\rm bg}(t),
\end{equation}
for $i=x,y,z$.

The monopole mode is then measured by the fluctuations
\begin{equation}\label{eq:mono}
\Delta \sigma_{\rm mono}^2 = (\Delta \sigma_x^2 + \Delta \sigma_y^2 + \Delta \sigma_z^2)/3
\end{equation}
while the quadrupole mode is measured by
\begin{equation}\label{eq:quad}
\Delta \sigma_{\rm quad}^2 = \Delta \sigma_x^2 - \Delta \sigma_y^2.
\end{equation}

These fluctuations are plotted in \fref{fig:mode_Ec1200} and \fref{fig:mode_Ec3200} for the condensate (top panels) and the thermal clouds (bottom panels). The oscillation amplitude is larger for the deeper cut because of the greater pressure that pushes the thermal atoms outwards at the start of the simulation (see the discussion near the end of section~\ref{sec:single_thermal}). We further determine the dominant oscillation frequencies by Fourier transforming the fluctuations~\eref{eq:mono} and \eref{eq:quad}.

In general, the monopole mode displays oscillation with angular frequency between $2\omega$ and $2.5\omega$, in reasonable agreement with the expected monopole frequency of a pure thermal cloud (2$\omega$)~\cite{guery-odeline_zambelli_1999} or a pure condensate ($\sqrt{5}\omega$)~\cite{stringari_1996}. The oscillation amplitude decays in time, but it is difficult to extract a decay time scale due to the lack of regular decaying pattern, associated with the underlying condensate growth.

Because of numerical fluctuations generated in our simulations, we can also see small amplitude quadrupole oscillation that displays more irregularity than the monopole oscillation. The fast Fourier-transformed fluctuations display peak between $\omega$ and $1.5\omega$, in reasonable agreement with the quadrupole mode frequency $\sqrt{2}\omega$ of a pure condensate~\cite{stringari_1996}.

Due to the complicated nature of the excitation mechanism of such processes, occurring on top of a constantly cooling sample undergoing condensate growth, we are unable at the present time to extract more information related to the induced in-phase and out-of-phase oscillations, which have proven crucial in interpreting experimental findings under controlled excitation schemes~\cite{jin_ensher_1996,olshanii_98,bijlsma_stoof_99,alkhawaja_stoof_00,jackson_zaremba_2002c,morgan_rusch_03,morgan_05}. A more detailed study of the coupled excitation amplitudes under different cooling protocols and geometries remains an interesting question for further studies.

\bibliography{JPhyB_2compo}

\end{document}